\documentclass{aa}  

\usepackage{graphicx}
\usepackage{txfonts}
\usepackage{amsmath}
\usepackage{amssymb}
\usepackage{natbib}
\usepackage{placeins}

\graphicspath{{./figures/}}
\bibpunct{(}{)}{;}{a}{}{,}
\newlength{\minuslength}
\begin{document}

   \title{A 3D physico-chemical model of a pre-stellar core.
I. Environmental and structural impact on the distribution of CH$_3$OH and $c$--C$_3$H$_2$.}


   \author{S. S. Jensen\inst{1}\thanks{\email{sigurdsj@mpe.mpg.de}} 
   \and S. Spezzano\inst{1}
   \and P. Caselli\inst{1}
   \and T. Grassi\inst{1}
   \and T. Haugb\o lle\inst{2}
   }

   \institute{Max-Planck-Institut f{\"u}r extraterrestrische Physik, Giessenbachstrasse 1, D-85748 Garching, Germany
   \and 
   Niels Bohr Institute, University of Copenhagen, {\O}ster Voldgade 5-7, DK-1350 Copenhagen K, Denmark}

   \date{Draft date: \today}

 
  \abstract
   {Pre-stellar cores represent the earliest stage of the star- and planet-formation process. By characterizing the physical and chemical structure of these cores we can establish the initial conditions for star and planet formation and determine to what degree the chemical composition of pre-stellar cores is inherited to the later stages.}
   {We aim to determine the underlying causes of spatial chemical segregation observed in pre-stellar cores and study the effects of the core structure and external environment on the chemical structure of pre-stellar cores.}
   {A 3D MHD model of a pre-stellar core embedded in a dynamic star-forming cloud is post-processed using sequentially continuum radiative transfer, a gas-grain chemical model, and a line-radiative transfer model. Results are analyzed and compared to observations of CH$_3$OH and $c$-C$_3$H$_2$ in L1544. Nine different chemical models are compared to the observations to determine which initial conditions are compatible with the observed chemical segregation in the prototypical pre-stellar core L1544.}
   {The model is able to reproduce several aspects of the observed chemical differentiation in L1544. Extended methanol emission is shifted towards colder and more shielded regions of the core envelope while $c$--C$_3$H$_2$ emission overlaps with the dust continuum, consistent with the observed chemical structure. These results are consistent across a broad spectrum of chemical models. Increasing the strength of the interstellar radiation field or the cosmic-ray ionization rate with respect to the typical values assumed in nearby star-forming regions leads to synthetic maps that are inconsistent with the observed chemical structure.}
   {Our model shows that the observed chemical dichotomy in L1544 can arise as a result of uneven illumination due to the asymmetrical structure of the 3D core and the environment within which the core has formed. This highlights the importance of the 3D structure at the core-cloud transition on the chemistry of pre-stellar cores. The reported effect likely affects later stages of the star- and planet-formation process through chemical inheritance.}

   \keywords{astrochemistry --
                radiative transfer -
                stars: formation --
                ISM: abundances --
                methods: numerical
                                   }

\titlerunning{Environmental impact on chemical structure of pre-stellar cores.}
\authorrunning{S. S. Jensen et al.}

   \maketitle
%
\section{Introduction}\label{sec:introduction}
Stars like the Sun form from the collapse of cold self-gravitating cores known as pre-stellar cores. The collapse of these cores marks the earliest stage of the star and planet formation process and consequently sets its initial conditions \citep[e.g.,][]{2007ARA&A..45..339B}. Studying the chemical inventory during this stage thus provides direct insights into the chemical inventory that will finally be inherited by forming planets. \citep[e.g.,][]{2012A&ARv..20...56C}. 

\begin{figure}[ht]
\resizebox{\hsize}{!}
        {\includegraphics{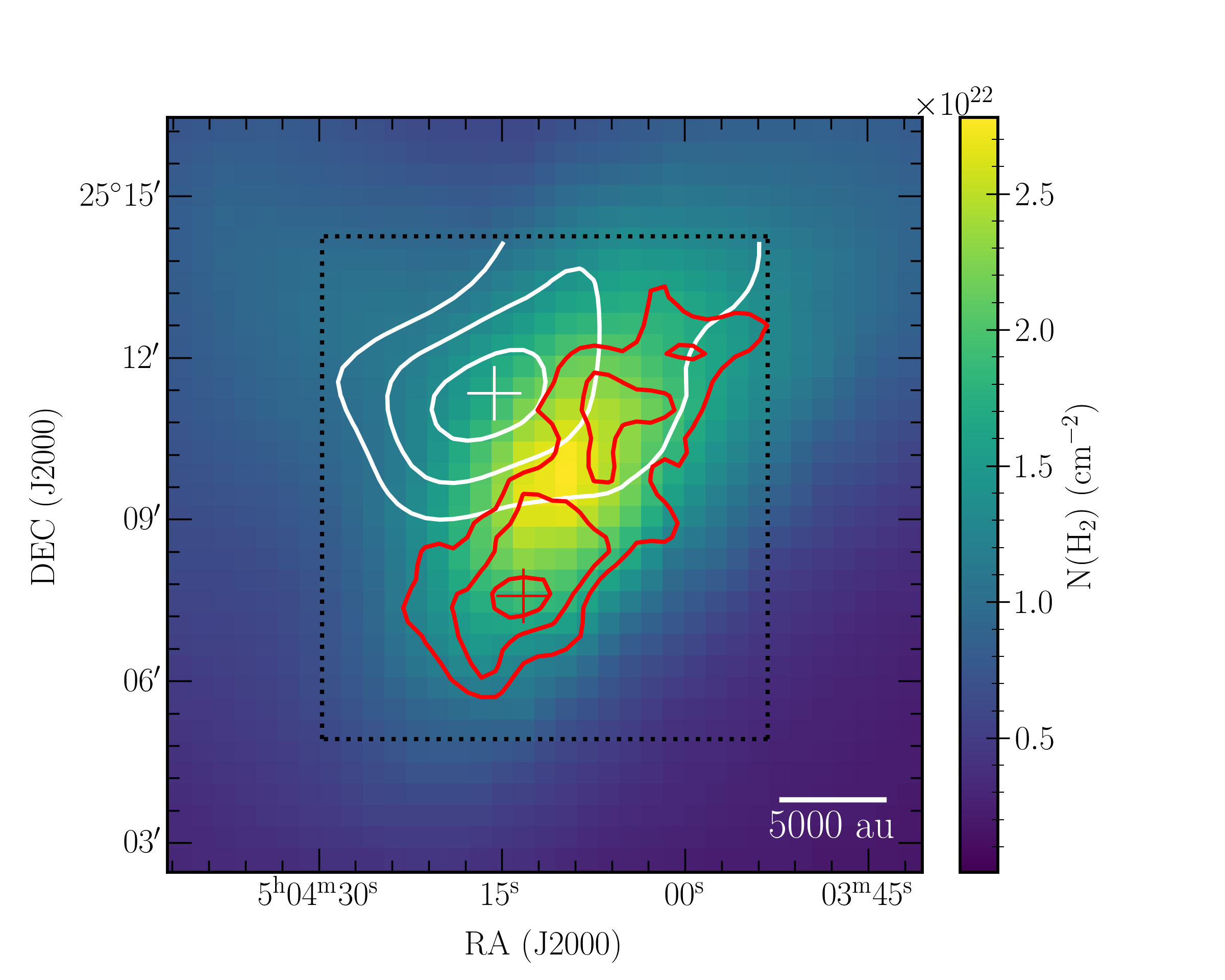}}
  \caption{Observations of $c$-C$_3$H$_2$ and CH$_3$OH toward L1544 with the IRAM 30m telescope. The red contours show the integrated emission of $c$-C$_3$H$_2$ 3$_{2,2}$--3$_{1,3}$ and the crosshair indicates the peak of the emission map. White contours and crosshair show the same for CH$_3$OH 2$_{1,2}$--1$_{1,1}$. For both molecules the contours indicates 50\%, 70\%, and 90\% of the peak emission. The black dotted square indicates the region mapped with the 30m. The colormap show the H$_2$ column density map derived from \emph{Herschel}/SPIRE observations. Both column density and molecular emission maps are from \citet{2016A&A...592L..11S}.}
     \label{fig:observations}
\end{figure}

Pre-stellar cores are found in a variety of different cloud environments, ranging from isolated Bok globules to dense clusters embedded in molecular clouds. How variations in the local environment influences the chemical composition of pre-stellar cores remains an open question.
In a recent work, \citet{2016A&A...592L..11S} mapped the emission of methanol (CH$_3$OH) and cyclopropenylidene ($c$-C$_3$H$_2$) in the pre-stellar core L1544 and found a complementary morphology in the emission of the two molecules as shown in Fig. \ref{fig:observations}. The chemical segregation observed in L1544 appeared to be caused by uneven illumination of the core by the interstellar radiation field (ISRF) because the methanol emission peaks on the north-western side of the core where the H$_2$ column density shows a more shallow drop than the remaining directions.
Later, \citet{2017A&A...606A..82S} mapped a large selection of molecules across L1544, and performed a principal component analysis on the emission maps. Their analysis confirmed the previous dichotomy between $c$-C$_3$H$_2$ and CH$_3$OH and found additional components linked to the dust peak of the core and HNCO. The analysis suggested that the emission morphology was determined by both local variations in temperature and the external variation in the incident ISRF. Furthermore, the observed dichotomy in $c$-C$_3$H$_2$ and CH$_3$OH has recently been reported in a larger sample of starless and pre-stellar cores \citep{2020A&A...643A..60S}. As such, the local cloud environment appears to influence the chemical structure of pre-stellar cores and may furthermore impact the chemical composition during planet formation through chemical inheritance.

There is growing evidence that the chemistry at the pre-stellar phase is partially inherited by the later stages of star and planet formation and consequently this environmental effect could influence the chemistry during planet formation. One prominent tracer of chemical inheritance is the degree of deuterium fractionation which record variations in the physical conditions during the formation of deuterated molecules \citep[e.g.][]{2014prpl.conf..835V, 2014prpl.conf..859C}. By comparing the D/H ratio in Solar System bodies with astrochemical models of the protosolar nebula, \citet{2014Sci...345.1590C} demonstrated that the observed D/H ratio in the Solar System cannot be formed in-situ and requires inheritance of pre-stellar water with a high D/H ratio. Likewise, the high D$_2$O/HDO ratios in embedded protostars indicate a substantial degree of inheritance of water from the pre-stellar phases \citep{2016A&A...586A.127F, 2021A&A...650A.172J}. Furthermore, comparison between the chemical content of the Class 0 binary source IRAS16293--2422 and Comet 67P provides tentative support for a high degree of inheritance for a larger number of molecules, as the relative abundances of complex organic molecules (COMs) show a correlation between the protostellar hot corino chemistry and the cometary composition \citep{2019MNRAS.490...50D}. These results suggest that the chemical composition during the pre-stellar phase plays a crucial role in the chemical composition in protoplanetary disks during the epoch of planet formation.

To understand the degree of chemical diversity among young star-forming systems, the effect of the different environments must be studied. Examples of previous work include the systematic studies between different molecular clouds \citep[e.g., PEACHES and ORANGES,][]{2021ApJ...910...20Y, 2022ApJ...929...10B} and comparisons between carbon-chain molecules and complex organic molecules in different regions of molecular clouds \citep[e.g.,][]{Higuchi_2018}. Another example is the comparison of the D/H ratio of water between isolated and clustered Class 0 protostars \citep{2019A&A...631A..25J}. These studies indicate a dichotomy between isolated and clustered protostars, and support an inheritance scenario for water. However, it is not yet understood what drives the apparent chemical differentiation between different environments and how strongly it can impact the chemical composition. 

In this work we aim to explore the origin of the observed chemical morphology of L1544 and establish what physical and chemical processes drive the chemical segregation in pre-stellar cores.
L1544 is a pre-stellar core located on the edge of the Taurus molecular cloud at a distance of 170~pc \citep{2018ApJ...859...33G}. Both the physical and chemical structure of the core has been the subject of several studies \citep[e.g.,][]{2002ApJ...569..815T, 2012ApJ...759L..37C, 2015MNRAS.446.3731K, 2019ApJ...874...89C, 2022ApJ...929...13C}. The center of L1544 is cold and dense, with the temperatures reaching 6.5 K and the volume density being as high as $10^7$ cm$^{-3}$  \citep{2007A&A...470..221C, 2019A&A...623A.118C}. Molecular line emission observed towards the dust peak of L1544 has been successfully modeled using the 1D physical model presented in \citet{2010MNRAS.402.1625K}. However, maps of molecular emission shows clear sign of a non-spherical structure \citep[e.g.,][]{2017A&A...606A..82S}. Both continuum emission and molecular emission shows an elongated structure, which is not accounted for in 1D models and a full 3D model is likely needed to explain the chemical segregation within the core. Furthermore, \citet{2019ApJ...874...89C} and \citet{2021MNRAS.505.5142Z} studied the collapse of a dense core using a 3D non-ideal MHD code and found that an asymmetric disk-like structure forms in the pre-stellar phase.
This work presents a 3D ideal-MHD model of a pre-stellar core analog to L1544. The model includes a detailed chemical network and realistic boundary conditions from a large-scale molecular cloud simulation. This allows us to study the chemical structure of a non-spherical pre-stellar core in a realistic dynamic star-forming environment and test the proposed origin of the chemical structure of the core. In this initial work we focus on the chemical structure of CH$_3$OH and $c$-C$_3$H$_2$, leaving other molecules for future publications.

The paper is organized as follows. The second section introduces each element of the model. In the third section, we compare both the continuum and the molecular emission from the simulated pre-stellar core with L1544. The implications of the model, along with caveats, are discussed in the fourth section and the conclusion are presented in the last section.

\section{Model description} \label{sec:methods}

\subsection{Physical pre-stellar core models}
\begin{figure*}[ht]
\resizebox{\hsize}{!}
        {\includegraphics{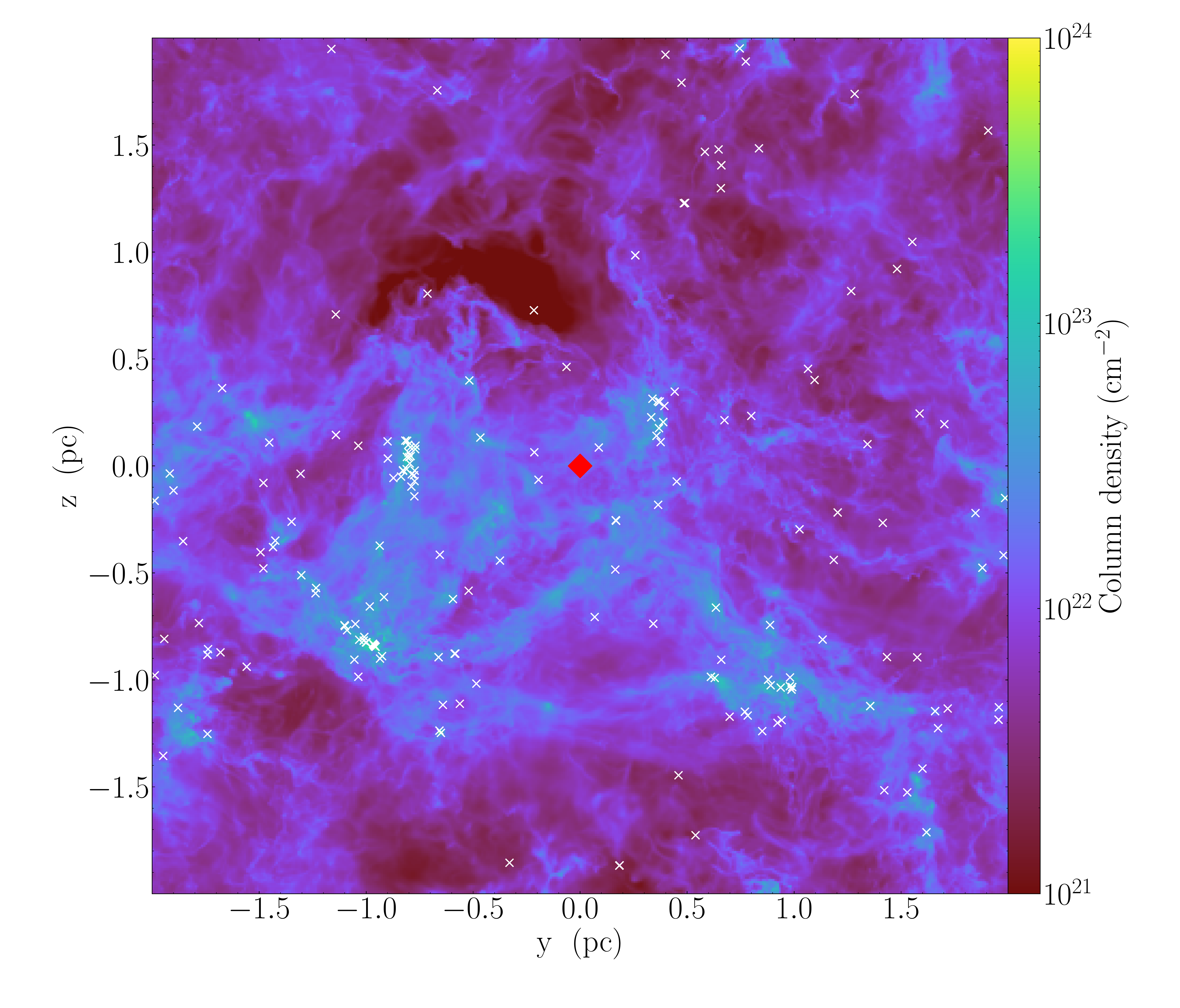}}
  \caption{Projected H$_2$ column densities in the molecular cloud simulation. A total of 233 protostars, denoted with white crosses have formed in this snapshot from the simulation (t = 1.66 Myr). Most protostars form in clusters along filaments. The red diamond marks the protostar which is studied here during the pre-stellar phase. We have centered the pre-stellar core in the figure for clarity, since the box has periodic boundary conditions.}
     \label{fig:ramses}
\end{figure*}

To study the impact of the surrounding star-forming environment, we selected a pre-stellar core embedded in a large-scale simulation of a 4 pc$^3$ star-forming molecular cloud. The dynamical evolution of the molecular cloud was modeled using the the adaptive mesh refinement finite volume ideal-MHD code {\sc ramses} \citep{2002A&A...385..337T}. 
A detailed description of the physical model is presented in \citet{2018ApJ...854...35H} and we will here briefly introduce the characteristic of the model.
In the model, turbulence in the gas is driven with a solenoidal powerspectrum on the largest scales maintaining a 3D rms velocity of 2 km~s$^-1$, and an isothermal equation of state is used. Sink particles that can accrete gas from the surroundings and represent stars form at the highest level of refinement, when a grid cell is detected as unequivocally collapsing \citep{2018ApJ...854...35H}. 134 million Lagranian tracer particles with 6.6 Earth masses each that passively follows the flow of the gas while recording density, velocities, and magnetic field strengths are used to record the flow of matter in the model. The molecular cloud model reproduce well the physical and statistical conditions of a local low mass star forming region including the star formation rate and initial mass function \citep{2018ApJ...854...35H}, core mass function \citep{2021MNRAS.504.1219P}, and the multiplicity distribution \citep{2022arXiv220901909K}.
The model is well suited to study star formation as a heterogeneous process, where the physical evolution on larger scales may impact the physical and chemical evolution on smaller scales.
The physical simulation uses adaptive mesh refinement (AMR) with a root grid of 512$^3$ and six levels of refinement. This results in a minimum cell-size of 25~au in models presented here. This resolution is sufficient to study the dynamics from the larger molecular cloud scales down to individual pre-stellar cores. Protostellar feedback, e.g., jets and outflows, were not included in the model, since we are concerned with pre-stellar phase in this work. In addition, L1544 is rather isolated so nearby protostars are unlikely to impact the structure of the core. The entire 4~pc$^3$ box is shown in Fig. \ref{fig:ramses}.

Within the computational domain we selected a pre-stellar core resembling L1544. A description of the selection criteria is presented in Sec. \ref{subsec:selection}.
Around the selected pre-stellar core we extracted a spherical cutout with a radius 5$\times10^4$~au which is used for the post-processing steps of the model, introduced in the subsequent sections. We carry out the post-processing and analysis on this reduced domain because performing a full radiative transfer simulation on the entire 4~pc$^3$ box is unfeasible given the number of photons needed to sample a domain of that size. The center of the core is defined as the cell with the highest density.

\subsection{Continuum radiative transfer}
To determine the thermal structure of the pre-stellar core, we computed the dust temperature in the core using the {\sc mctherm} routine from the Monte Carlo radiative transfer code {\sc radmc-3d} \citep{2012ascl.soft02015D}. The radiation source was the ISRF, irradiating the core from the edges of the 5$\times10^4$~au cutout around the pre-stellar core. A high number of photons was necessary to sample the central parts of the core due to the high opacity and large computational domain and we chose to inject 2.5$\times10^8$ photons in this work.
We used the ISRF from \citet{2001A&A...376..650Z}, with the modification in the mid-infrared as presented in \citet{2017A&A...604A..58H}. The UV range of the ISRF is composed of the Draine field \citep{1978ApJS...36..595D}, similar to the one used to compute the photo-rates in \citet{2017A&A...602A.105H} for consistency with the chemical model. We denote the standard strength of the ISRF as $G$. 
In the fiducial model, we assumed an external extinction of A$_{\mathrm{v, ext}}$ = 2 mag around the core. This value was previously used in the 1D radiative hydrodynamical model presented in \citet{2017A&A...607A..26S}. Values with A$_{\mathrm{v, ext}}$ = 0 mag and A$_{\mathrm{v, ext}}$ = 4 mag were also tested. A lower limit on the temperature of 6~K is imposed. Due to the high density in the center of the core and the large computational domain, some cells will not be well sampled (i.e., low photon counts) and have temperatures below this limit. We do not consider this to be physical but rather a limitation of the current model. Nonetheless, the hard limit of 6~K does not have a notable impact on the results presented here as the cells with temperature below 6~K represent less than 0.6\% of the total number of cells.

We assumed that dust and gas are well-coupled and remain in thermal equilibrium. This assumption only holds at higher densities \citep[$\gtrsim10^5$ cm$^{-3}$,][]{2001ApJ...557..736G}, however previous 1D studies of pre-stellar cores with coupled radiative hydrodynamics have found that the difference between gas and dust temperatures remains limited \citep[i.e., $\Delta T\lesssim$~2~K,][]{2017A&A...607A..26S} and consequently the impact on the chemistry is negligible compared to the uncertainties on other parameters in grain-surface chemical models.

The fiducial model used the dust opacities for protostellar cores presented in \citet{1994A&A...291..943O} for thin ice layers and dust coagulation for 10$^5$ years at a density of n = 10$^6$~cm$^{-3}$ (from here-on referred to as OH94). These opacities provide a good match to observations of pre- and protostellar cores \citep[e.g.,][]{1999ApJ...522..991V, 2001ApJ...557..193E, 2002A&A...389..908J}. However, using a single set of dust opacities across the core is a simplification, since the ice thickness, composition, and grain size may vary across the core, impacting the opacity of the dust grains. Changing the dust opacity to bare dust grains or grains covered with thick ice layers from \citet{1994A&A...291..943O} have only limited impact on the thermal structure of the core presented here. 

In addition to the thermal dust continuum calculation, we calculated the radiation field in the core using the {\sc mcmono} routine in {\sc radmc3d} for a single wavelength $\lambda = 550$~nm. This allows us to estimate the visual extinction in each cell of the model, which was used for the chemical post-processing.
We compute the visual extinction following the approach of \citet{2017A&A...604A..58H}:
\begin{equation}
    A_\mathrm{v} = 1.086 \times \tau_{550} ~,
\end{equation}
where $\tau_{550}$ is the optical depth at 550~nm computed from:
\begin{equation}
    \tau_{550} = - \log \frac{I_\mathrm{cell,550}}{I_\mathrm{ISRF,550}} ~,
\end{equation}
Here $I_\mathrm{cell,550}$ denotes the computed intensity in a cell at 550~nm while $I_\mathrm{ISRF,550}$ is the unattenuated intensity of the interstellar radiation field at the same wavelength.

\subsection{Chemical model}
The chemical structure of the pre-stellar core was computed using a rate-equation approach \citep{1992ApJS...82..167H}. Rather than calculating the evolution independently for each cell (> 10$^6$ cells), we computed the chemical evolution for a grid of physical conditions and used linear interpolation to estimate the chemical composition in each cell. The interpolation grid consisted of six dimensions: total density of hydrogen $n_H$, extinction $A_\mathrm{v}$, temperature $T$, ISRF factor $G$, and cosmic-ray ionization rate $\eta_\mathrm{CR}$. A linear interpolation was performed using the {\sc NDLinearInterpolation} routine from {\sc scipy} \citep{2020SciPy-NMeth}, which relies on {\sc qhull} \citep{10.1145/235815.235821} for interpolation in higher dimensions. Density and cosmic-ray ionization rate were interpolated in log-space. To validate the interpolation method, we compared direct and interpolated chemical abundances on the 1D model of L1544 derived in \citet{2015MNRAS.446.3731K} in Appendix \ref{app:0}.

The chemical model was introduced in \citet{2021A&A...649A..66J} and we will limit this introduction to the key aspects.
The gas-phase network was based on the {\sc kida} network \citep{2015ApJS..217...20W}, extended to include deuterated species and spin-chemistry for H$_2$ and H$_{3}^{+}$ as presented in \citet{2017MNRAS.466.4470M} and \citet{2017A&A...607A..26S}. A total of $\sim$950 gas-phase species and $\sim$50000 reactions are included. Furthermore, a number of photo dissociation rates have been updated based on \citet{2017A&A...602A.105H}. Three-body reactions were not included in the network.

The grain surface network was based on the OSU network \citep[e.g.,][]{2007A&A...467.1103G} and includes $\sim$450 species and $\sim$3500 reactions. The grain-surface formation pathway for methanol formation is adapted from \citet{2014ApJ...791....1T}. Deuterated isotopologs are included for species with up to four carbon atoms. Surface reactions occur through the Langmuir-Hinshelwood mechanism using thermal diffusion on the grain surface. The chemical model considered a single grain size, with a radius of $10^{-5}$~cm and 1.8$\times10^{6}$ sites per grain \citep[e.g.,][]{2010A&A...522A..42S}. Reaction barriers can either be overcome thermally or through quantum tunneling, depending on which is faster. A barrier width of 1.0 \AA~was assumed except for the reactions listed in Appendix \ref{app:reactions}. The model included reaction-diffusion competition as introduced in \citet{2007A&A...469..973C}. The effect of grain coagulation on the chemistry was not considered in this work.

The grain-surface network included a number of non-thermal desorption mechanisms which may be crucial in starless and pre-stellar cores. Chemical desorption was included following the implementation presented in \citet{2007A&A...467.1103G} with an efficiency factor of 0.01. 
Cosmic-ray (CR) desorption was included using the description from \citet{1985A&A...144..147L} and \citet{1993MNRAS.261...83H}. An updated treatment of CR desorption is presented in \citet{2021ApJ...922..126S}, but is not yet implemented here. We consider both photodesorption by CR-induced UV photons and UV photons from the ISRF.
Photodesorption yields for N$_2$, CO, CO$_2$, and H$_2$O were based on laboratory studies of \citet{oberg2007, oberg2009} and \citet{fayolle2011, fayolle2013}. For the remaining species, a fixed desorption yield of $10^{-3}$ was adopted \citep{2015A&A...584A.124F}.
The network included photodissociation of grain-surface species. The rates here were approximated using the approach from \citet{2015A&A...584A.124F} which estimates the grain-surface photodissociation rates by rescaling the gas-phase photodissociation rates. The scaling factor was derived from the ratio of the H$_2$O photodissociation rates in gas and on grain surfaces, which had been calculated explicitly. The full network is available in the chemical model repository \footnote{\url{https://github.com/ssjensen92/kemimo}}.

In \citet{2021A&A...649A..66J} a three-phase model was utilized (i.e., separation of surface and mantle chemistry) while a simpler two-phase model was utilized here. This was done to reduce computational time. Furthermore, \citet{2017A&A...607A..26S} found that their two-phase model provides a better match to observations of L1544 than their three-phase model for a subset of deuterated species. We therefore opted for the simpler model to reduce the impact of uncertain parameters such as binding energies for surface and mantle species and reduce the overall complexity of the chemical model. Exploring the differences between the two and three phases models goes beyond the scope of the present paper.

\begin{table}[ht]
\caption{Initial abundances following \citet{2010A&A...522A..42S}, except for the inclusion of ortho and para spin-states of H$_2$. The initial ortho-to-para ratio is set to 10$^{-3}$.}
\begin{tabular}{ll|ll}
\hline
\hline
Species & Abundance (/$n_\mathrm{H}$) & Species & Abundance (/$n_\mathrm{H}$) \\ \hline
o-H$_2$ &      4.995(-1)                  & S$^{+}$      & 8.00(-8)                    \\ 
p-H$_2$ &      5.00(-4)                  & Si$^{+}$     & 8.00(-9)                        \\ 
HD      &      1.60(-5)                  &         Mg$^{+}$     & 7.00(-9)                \\ 
He      &      9.00(-2)                  & Fe$^{+}$     & 3.00(-9)                    \\ 
O       &      2.56(-4)                 & Na$^{+}$     & 2.25(-9)                       \\ 
C$^{+}$       &      1.20(-4)                     &  Cl$^{+}$     & 1.00(-9)                       \\ 
N       &      7.60(-5)              &         P$^{+}$      & 2.00(-10)                \\ 
\hline
\end{tabular}
\label{tab:abundances}
$A(B) = A\times10^{B}$.
\end{table}

Our fiducial model starts from the elemental abundances listed in Table \ref{tab:abundances} and is evolved for 10$^6$ years for each point in the grid. A standard ISRF intensity $G_0 = 1$ is used together with a cosmic-ray ionisation rate of 3$\times10^{-17} \mathrm{s}^{-1}$. This value was recently proposed as the best-fit constant value for L1544, based on 1D astrochemical modeling of ionic emission fitted to observations \citep{2021A&A...656A.109R}.
Since the initial conditions and the timescale of the chemical model can have a substantial impact on the chemical abundances we tested a number of alternative chemical models. Variations include the addition of a low-density initial phase $T_\mathrm{init}$ at a fixed density and temperature ($n_\mathbf{H} = 5\times10^2\mathrm{cm}^{-3}$, $T = 15$~K), different chemical timescales $T_\mathrm{main}$ in the entire physical grid and variations in the intensity of the ISRF and the cosmic-ray ionization rate.  An overview is listed in Table \ref{tab:chemical_models}.

\begin{table*}
\caption{Overview of the different chemical setups included in this work. The parameter $T_\mathrm{init}$ indicates the duration of a static initial phase at $T = 15$~K, $n_\mathbf{H} = 5\times10^2\mathrm{cm}^{-3}$, and $A_\mathrm{V}$ = 2 mag. The fiducial model is labeled $\mathrm{I0.0\_M1.0\_standard}$.}\label{tab:chemical_models}
\centering
\begin{tabular}{llllll}
\hline\hline
Identifier & $T_\mathrm{init}$ (Myr) & $T_\mathrm{main}$ (Myr) & $\zeta_\mathrm{CR}$ (10$^{-17}$ s$^{-1}$) & $G_0$ & comment \\ \hline
$\mathrm{I0.0\_M1.0\_standard}$   & 0.0                  & 1.0                      & 3        & 1 & \emph{Fiducial}        \\
$\mathrm{I0.0\_M0.5\_standard}$    & 0.0                  & 0.5                      & 3   & 1     &         \\
$\mathrm{I0.5\_M0.2\_standard}$   & 0.5                  & 0.2                      & 3   & 1     &         \\
$\mathrm{I0.5\_M0.5\_standard}$   & 0.5                  & 0.5                      & 3    & 1    &        \\
$\mathrm{I1.0\_M0.1\_standard}$   & 1.0                  & 0.1                      & 3    & 1    &   \\
$\mathrm{I0.0\_M1.0\_}\kappa\times0.2$  & 0.0                  & 1.0                      & 3        & 1 &$\kappa_\mathrm{dust} = 0.2\times \kappa_\mathrm{OH94}$     \\
$\mathrm{I0.0\_M1.0\_Gnot100}$   & 0.0                  & 1.0                      & 3    & 10$^2$    &        \\
$\mathrm{I0.0\_M1.0\_CR10}$   & 0.0                  & 1.0                      & 30    & 1    &        \\
$\mathrm{I0.0\_M1.0\_extAv0}$   & 0.0                  & 1.0                      & 3    & 1    &     A$_\mathrm{V,ext} = 0$   \\
\hline
\end{tabular}
\end{table*}

\subsection{Line radiative transfer}
The last step of the model is a non-LTE calculation of molecular emission from selected molecules. We used the 3D {\sc LIME} radiative transfer code \citep{2010A&A...523A..25B}. {\sc LIME} is a Monte Carlo radiative transfer code which solves the radiative transfer equation using the accelerated lambda iteration (ALI) method \citep{1991A&A...245..171R}. The computational grid in {\sc LIME} was constructed using 3$\times10^{4}$ randomly sampled points $P$ and the domain is limited to a radius of $2.5\times10^{4}$~au. We limit the domain to this region to reduce computational time. The reduced domain matches the region that is usually mapped across L1544 with the 30m telescope \citep[e.g.,][]{1999ApJ...523L.165C,2016A&A...592L..11S, 2019A&A...629A..15R}. 
The code connects the random point distribution by performing a Delauney triangulation in 3D using the Qhull library \citep{10.1145/235815.235821}. The grid is then constructed by performing Voronoi tesselation. In Voronoi tesselation, individual cells are created for each point $P$. The extent of each Voronoi cell is defined as the 3D region nearest to a point $P$. For a complete description of the grid construction we refer to \cite{2010A&A...523A..25B}.
The molecular level populations based on collisional rates is provided in the LAMDA format \citep{2005A&A...432..369S}. We ran 24 iterations for all molecules, which was sufficient to insure convergence of the level populations.
LIME computes a spectral cube of the simulated frequency range from which both molecular maps and spectra can be extracted. We place our source at a distance of 170~pc, similar to the distance to L1544  \citep{2018ApJ...859...33G}, and convolve the spectral cubes with a 2D Gaussian beam to match the observations which we will compare with the simulated images in the subsequent sections.

\section{Results} \label{sec:results}
\subsection{Selecting a L1544 candidate} \label{subsec:selection}
We selected a pre-stellar core from the {\sc RAMSES} simulation which resembles L1544 both in terms of its intrinsic characteristics such as mass and central density and the surrounding environment. The estimated mass of L1544 varies based on the method employed. From C$^{18}$O (1--0) observations, \citet{1998ApJ...504..900T} estimated a mass of 8~$\mathrm{M}_\odot$ within a contour of approximately 0.1x0.2~pc. However, estimates from continuum observations point to a lower mass of $\sim$2~$\mathrm{M}_\odot$ \citep[e.g.,][]{2001ApJ...557..193E}. This difference of a factor of four likely stems from the assumptions and limitations of each method. 
The molecular estimate assumes a fixed relative abundance $X_\mathrm{CO}$ across the core, not accounting for CO freeze-out. Meanwhile, mass estimates derived from continuum observations assume a single dust opacity across the core and are derived from observations using the 'chopping' method. The use of chopping results in some filtering of extended envelope emissions which contribute to the mass estimated using molecular tracers \citep{2000ApJS..131..249S}. The assumption of a single dust opacity is not likely to be satisfied. At lower densities in the exterior shells of the core, dust grains are likely covered by less ice while dust grains at higher densities are covered by thick ice shells and may undergo coagulation. Furthermore, changes in the ice composition, e.g., due to CO freeze-out, may also impact the exact opacity.
Both mass estimates also assume a single temperature across the core. To summarize, the mass of L1544 is not well-constrained, but estimates lie in the range $M \in [2.0, 10.0]~\mathrm{M}_\odot$. We selected a core with a mass of 8.25~M$_\odot$ within a radius of 2$\times10^4$~au, compatible with the estimated mass range.

Recent interferometric observations have revealed a kernel at the center of the core, with a radius of 1800~au and a mean density of 10$^6$ particles cm$^{-3}$ \citep{2019ApJ...874...89C}. The selected core in this study has a mean density of 1.2$\times10^6$ particles within a radius of 1800~au. The line-of-sight CO depletion factor in L1544 was estimated by \citet{1999ApJ...523L.165C} to be of the order $\sim$10 using single-dish observations of $^{17}$CO (1--0) and continuum observations while the core studied here has a line-of-sight CO depletion factor of $\sim$50 on similar scales (i.e., within $r \sim$6500~au). This corresponds to a local CO depletion factor of more than 1000 in the central region of the core.

\begin{figure*}[ht]
\resizebox{\hsize}{!}
        {\includegraphics{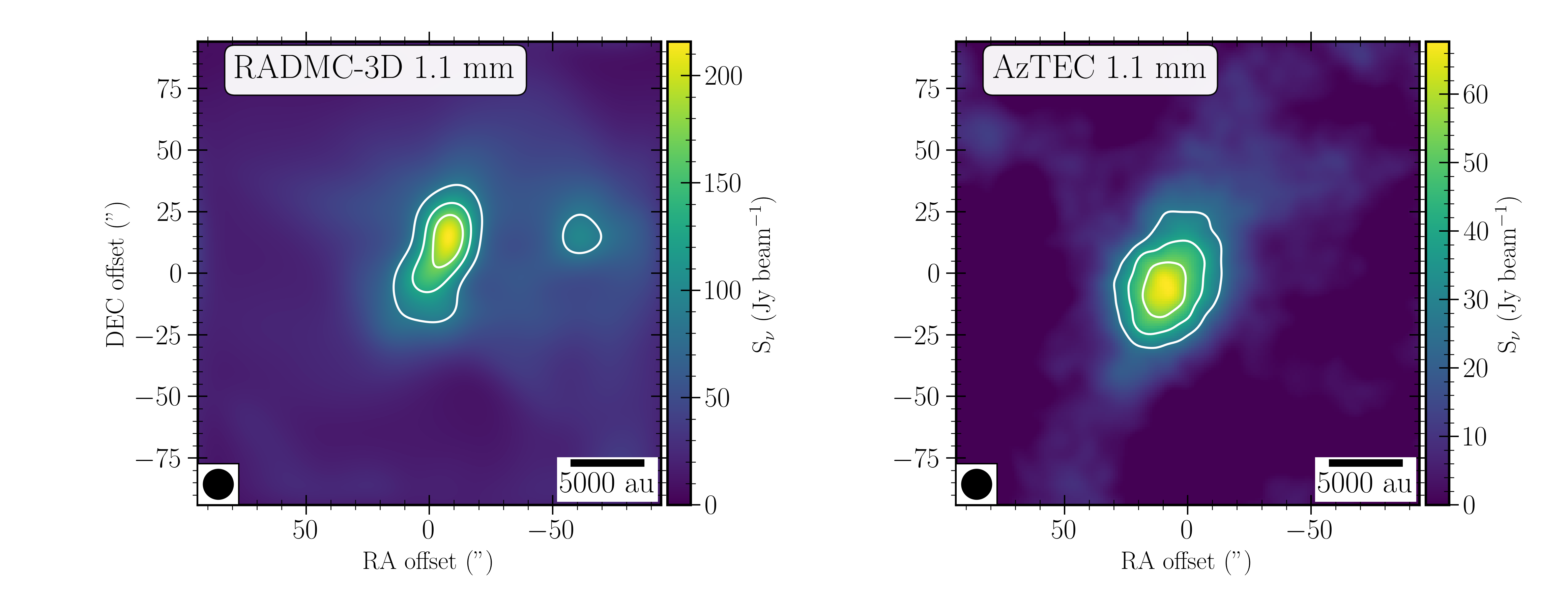}}
  \caption{Comparison of the continuum at 1.1 mm as modeled using {\sc radmc-3d} and as observed using the AzTEC continuum camera at the LMT \citep[presented in ][]{2019A&A...623A.118C}. Contours indicate 40\%, 60\%, and 80\% of the peak intensity in each case. The beam-size of the AzTEC bolometer is 12.6~$\!\!^{\prime\prime}$ and the synthetic observation has been convolved with a Gaussian of the same size. The linear scale is shown in the lower left of each panel, assuming a distance of 170~pc to L1544 and the model core. The beam size is shown in black in the lower left corner.}
     \label{fig:cont_comparison}
\end{figure*}

\begin{figure}[]
\resizebox{\hsize}{!}
        {\includegraphics{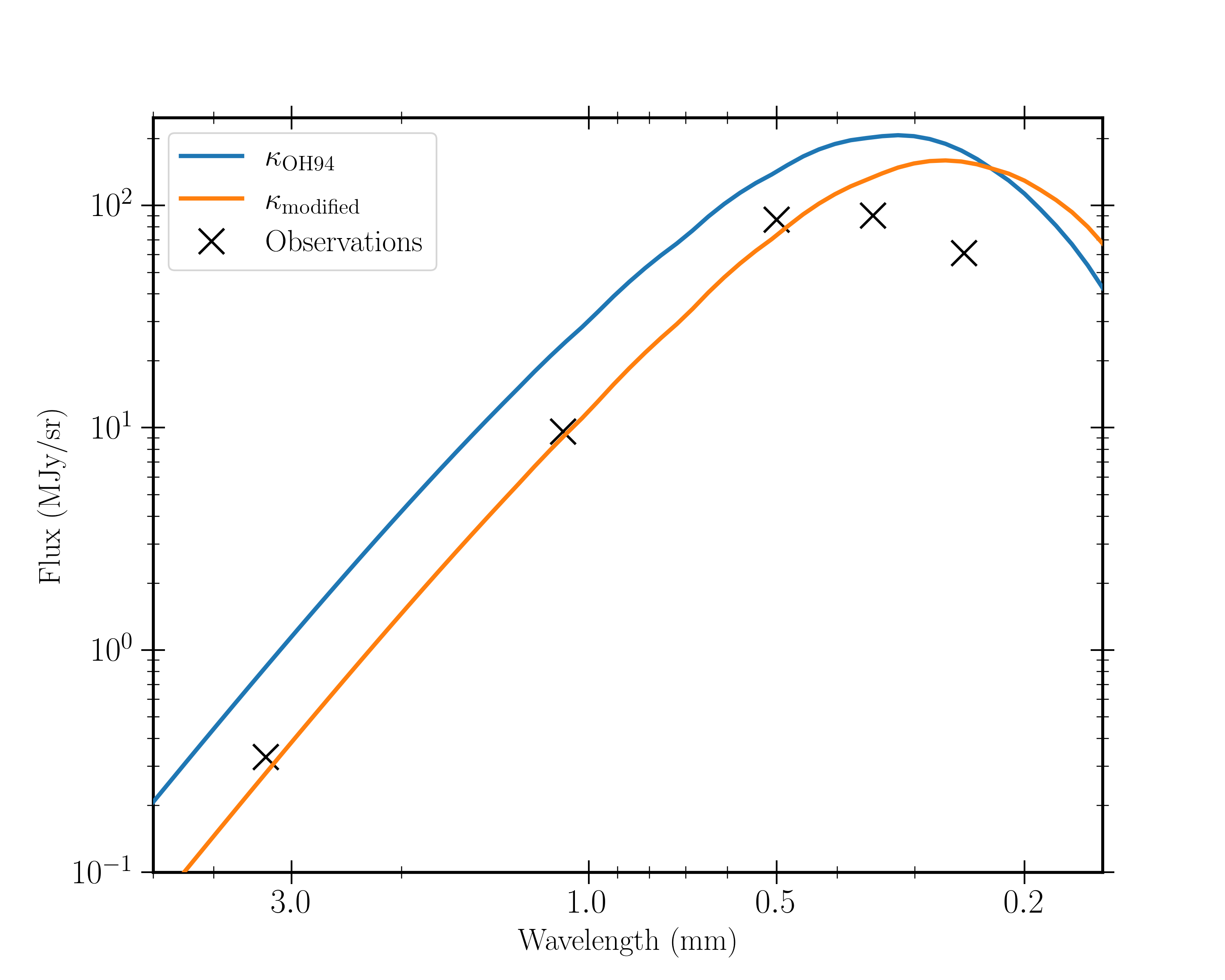}}
  \caption{Comparison between the spectral energy distribution of the modeled core and L1544. The observed values for L1544 are indicated with black crosses and are taken from the work by \citet{2017A&A...606A.142C}. The blue line shows the SED for the core model computed using the standard OH94 dust opacities, while the orange line shows the SED when the dust opacities have been multiplied by a factor of 0.2 to match the continuum emission at 1.1~mm.}
     \label{fig:SED}
\end{figure}

In terms of the local cloud environment, L1544 is located at the edge of the Taurus molecular cloud and shows evidence of uneven illumination \citep{2016A&A...592L..11S}. We therefore searched for a pre-stellar core formed in relative isolation (i.e., no nearby stars during the pre-stellar phase). The selected pre-stellar core is not representative of a Bok globule \citep[i.e., not like B68, ][]{2001Natur.409..159A} as it is still embedded in the parent molecular cloud. The core is isolated during the pre-stellar phase and remains so for the first 2$\times10^5$ years after the formation of the protostar.

A simulated continuum image of the selected L1544 analog is shown in Fig. \ref{fig:cont_comparison} along side continuum observations presented in \citet{2019A&A...623A.118C}. The spectral energy distribution (SED) of the core is shown in Fig \ref{fig:SED} and compared with the SED presented in \citet{2017A&A...606A.142C}. As found in previous works \citep[e.g.,][]{2001A&A...376..650Z}, the OH94 dust opacities lead to a brighter continuum than what is observed, which is also evident from the SED in Fig. \ref{fig:SED}. A lower dust opacity of $0.2 \times \kappa_\mathrm{OH94}$ is also included for comparison. A lower-mass core would bring the modeled continuum closer to the observed value, but be inconsistent with the derived molecular masses. A more realistic dust model, accounting for different dust sizes and ice coverage and the consequent variations in opacity would likely be needed to remove these discrepancies.

The morphology of the selected core shares similarities with L1544. The core is not spherically symmetric, but rather elongated along the south-west to north-east diagonal in the sky. The flattened disk-like appearance can arise from either the compressive flows in the turbulent molecular cloud or due to the preferential collapse of the cloud parallel to the direction of the magnetic field \citep[e.g.,][]{2000ApJ...529..925C}.

\subsection{Comparing 1D and 3D models of L1544}
The 1D spherically symmetric model presented in \citet{2010MNRAS.402.1625K} has been used in a number of works to successfully reproduce line profiles of several molecules toward the dust peak \citep[e.g.,][]{2012ApJ...759L..37C, 2019A&A...629A..15R, 2022A&A...664A.119G}. However, the dust continuum emission toward L1544 shows a clear asymmetry which may impact the chemical structure on larger scales. In this section, we first compare the prestellar core structure from the 3D MHD simulation to the 1D hydrodynamical structure from \citet{2010MNRAS.402.1625K}. Later, we compare the radial abundance profiles of select species.

\begin{figure*}[ht]
\resizebox{\hsize}{!}
        {\includegraphics{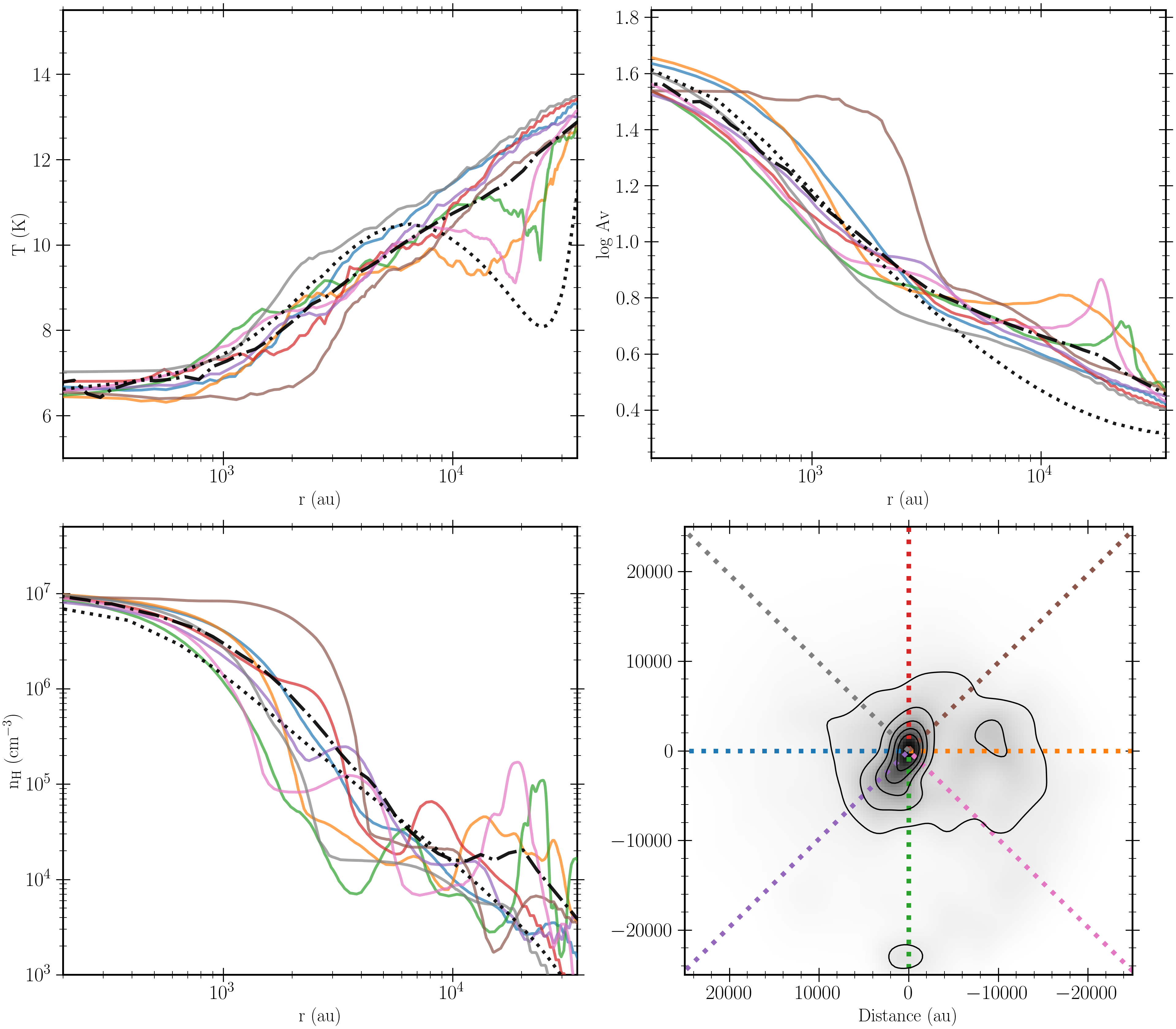}}
  \caption{The physical structure of the simulated core in the x-y plane. \emph{Left and top right}: Radial profile for the dust temperature, number density and visual extinction are shown along 8 different radial cuts on the x-y plane for the simulated core. The profiles for the 1D hydrodynamical model from \citet{2010MNRAS.402.1625K} is indicated by the black dotted line while the dashed-dotted line shows the mean of the 8 profiles in the 3D model. \emph{Lower right}: The greyscale images shows the continuum image of the core at 1.1~mm, while contours indicate 15\%, 30\%, 45\%, 60\%, and 75 \% of the peak intensity. The dashed colored lines indicate the direction along which the radial profiles in the left panel were extracted.}
     \label{fig:av_1d_3d}
\end{figure*}

\begin{figure}[ht]
\resizebox{\hsize}{!}
        {\includegraphics{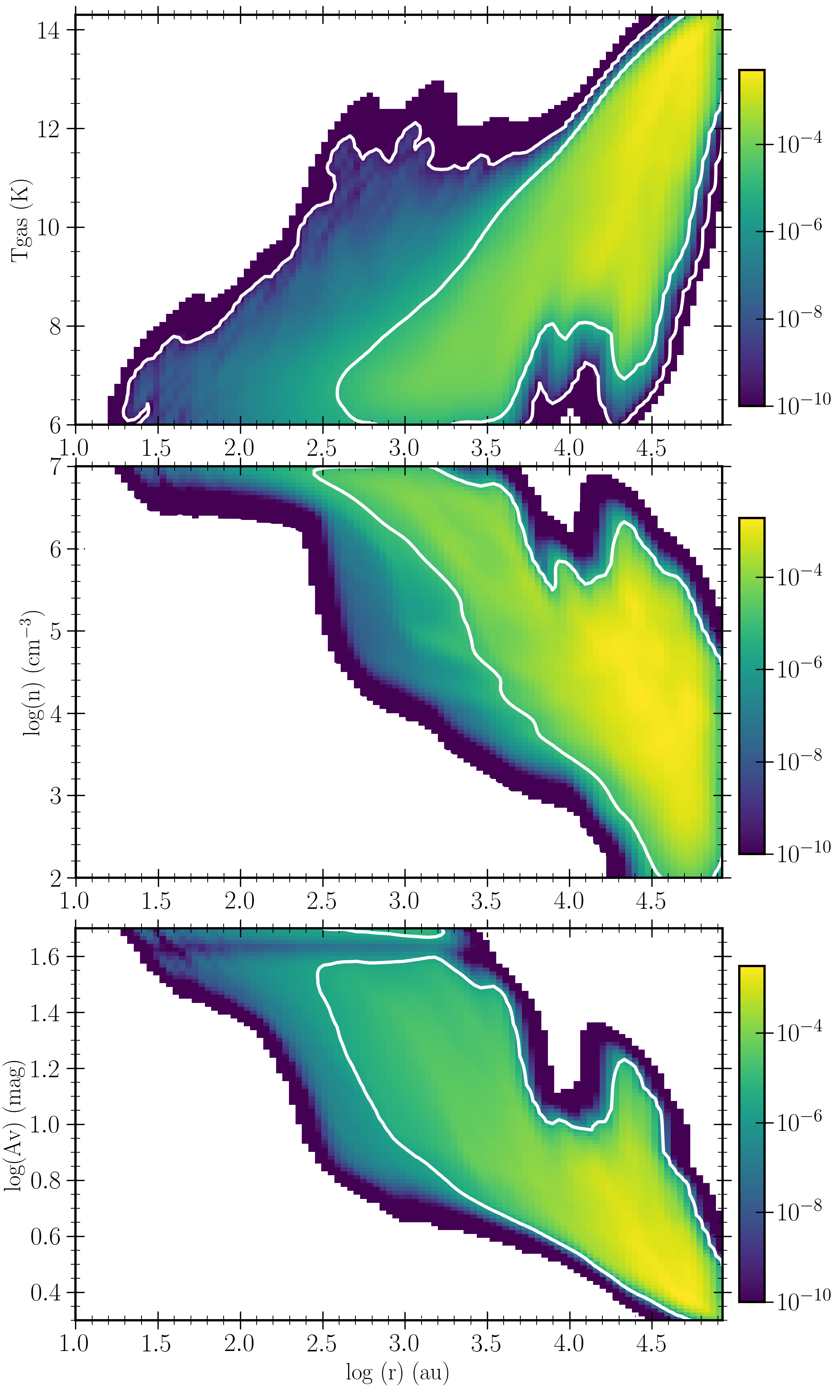}}
  \caption{Gaussian KDE for the cell temperature, density and visual extinction as a function of radius. White contours indicate the 50th and 75th percentiles.}
     \label{fig:KDE_3x1}
\end{figure}

In Fig. \ref{fig:av_1d_3d} we show a comparison of the radial temperature, density, and visual extinction in the 3D MHD model and the canonical 1D model from \citet{2010MNRAS.402.1625K}. The 3D model shows higher temperatures for the region with $r > 10^3$~au. The variation in temperature stems from both variations in the assumed dust opacity, but also variation in the large-scale structure of the core. For the density profiles, the 1D model resembles a mean profile of the 3D profiles out to around 10$^4$~au. The visual extinction profiles for the 3D model show some deviations from the 1D profile, especially at larger radii (e.g., the pink and orange profiles). Most notably is the brown profile for which the visual extinction increases beyond A$_\mathrm{v}$ = 10 at a radius of $\gtrsim3\times10^3$~au, while the remaining profiles reach A$_\mathrm{v}$ = 10 at $r < 2\times10^3$~au. The higher shielding can locally increase the CO freeze-out and promote the formation of complex organic molecules \citep{2017ApJ...842...33V}.
Overall, the 3D model clearly shows a deviation from the spherical model. Nonetheless, the structure of the 1D model is a good match to the 3D model at lower radius, where the density is high, and also follows a similar trend as the average profile for the 3D model \citep{2000ApJ...529..925C, 2019A&A...623A.118C}. To provide a visual indication of the asymmetries in the 3D model, we have computed Gaussian kernel density estimates (KDE) for the cell temperature, density and extinction as a function of radius. The results are shown in Fig. \ref{fig:KDE_3x1} and show that the physical conditions present a large spread due to the asymmetrical core structure. At 1000~au, the local density varies by several orders of magnitude (10$^{5}$--10$^{7}$ cm$^{-3}$), the temperature varies between 6--9~K, and the visual extinction ranges from ~7--35~mag.

In Appendix \ref{app:2} we have included the radial profiles similar to Fig. \ref{fig:av_1d_3d} for the remaining two coordinate planes of the model.

\subsection{Chemistry of $c$-C$_3$H$_2$ and CH$_3$OH}
\begin{figure}[ht]
\resizebox{\hsize}{!}
        {\includegraphics{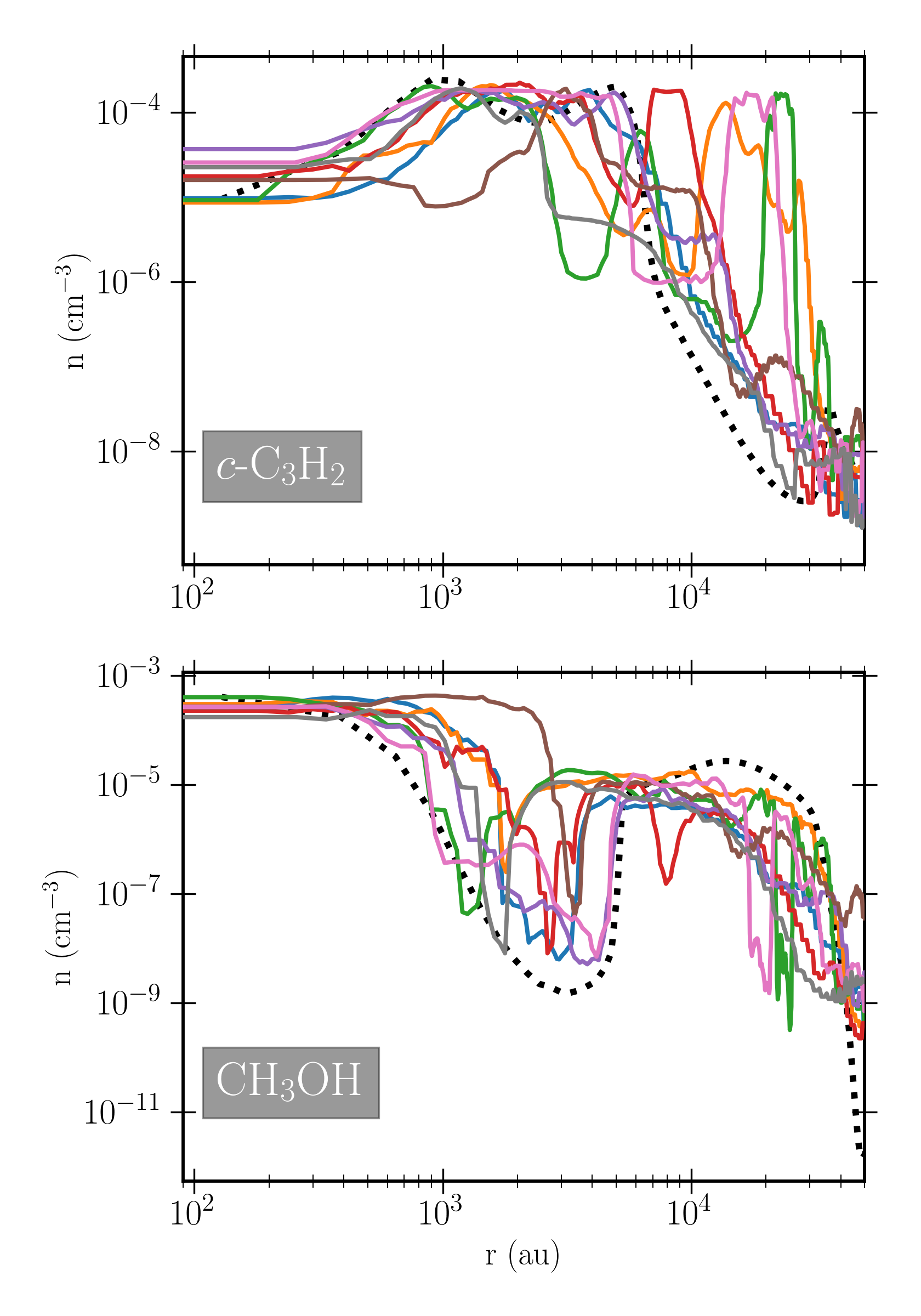}}
  \caption{Radial abundance profiles for $c$-C$_3$H$_2$ (top) and CH$_3$OH (bottom) along the 8 different profiles matching the colors shown in Fig. \ref{fig:av_1d_3d}. The profiles for the 1D hydrodynamical model from \citet{2010MNRAS.402.1625K} is indicated by the black dotted line.}
     \label{fig:c_C3H2_1D_3D}
\end{figure}

\begin{figure}[ht]
\resizebox{\hsize}{!}
        {\includegraphics{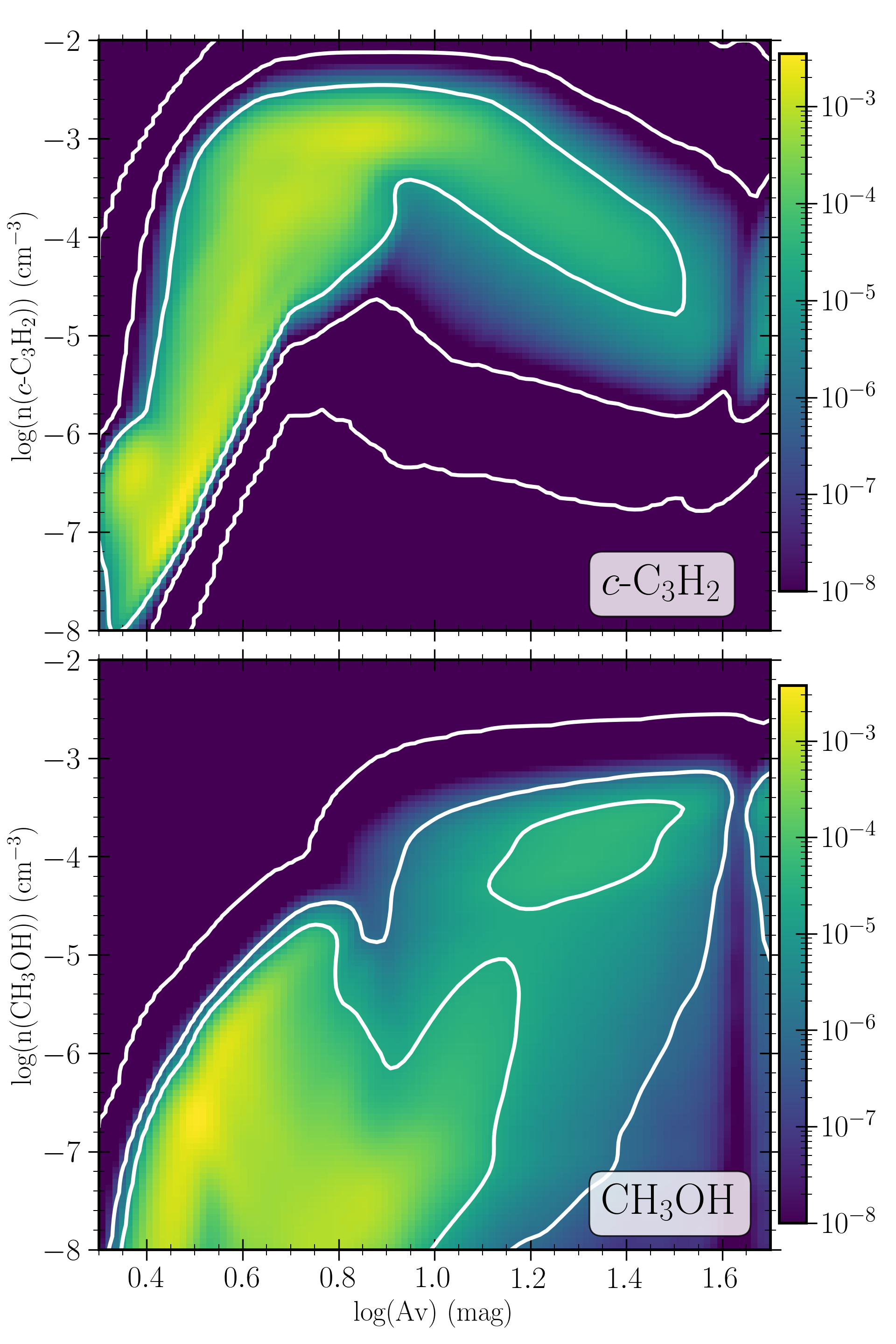}}
  \caption{Kernel density estimate for the chemical abundances as a function of the visual extinction. The analysis is carried out on a cell-by-cell basis in the 3D model and weighted by cell mass. Contours show the 25th, 50th, and 75th percentiles. Top panel shows $c$-C$_3$H$_2$ and bottom panel shows CH$_3$OH.}
     \label{fig:c_C3H2_kde}
\end{figure}

The formation of methanol in star-forming regions has been studied extensively and occurs primarily through sequential hydrogenation of CO on grain surfaces \citep[e.g., ][]{2002ApJ...571L.173W,2009ApJ...702..291H, 2009A&A...505..629F, 2016A&A...585A..24M}: 
\begin{equation}\label{eq:hydrogenation}
    \mathrm{CO} \xrightarrow{\mathrm{H}} \mathrm{HCO} \xrightarrow{\mathrm{H}} \mathrm{H}_2\mathrm{CO} \xrightarrow{\mathrm{H}} \mathrm{CH}_3\mathrm{O} \xrightarrow{\mathrm{H}} \mathrm{CH}_3\mathrm{OH}~.
\end{equation}
Thus, the formation of methanol is dependent on the presence of CO ice and H atoms on the dust grains, both of which dependent strongly on the local temperature, irradiation, and density. Indeed, the formation of methanol is especially efficient at the catastrophic CO freeze-out zone of pre-stellar cores \citep{1999ApJ...523L.165C, 2017ApJ...842...33V}, as CO ice allow CH$_3$OH molecules to return in the gas phase via reactive desorption more efficiently than H$_2$O ice. The formation of $c$-C$_3$H$_2$ on the other hand occurs through dissociative recombination in the gas-phase \citep[e.g., ][]{1985ApJ...299L..63T, 2016A&A...591L...1S}:
\begin{equation}\label{eq:cyclo1}
    \mathrm{C}_{3}^{}\mathrm{H}^{+} + \mathrm{H}_2  \rightarrow c\mathrm{-C}_{3}^{}\mathrm{H}_{3}^{+}~,
\end{equation}
\begin{equation}\label{eq:cyclo2}
    c\mathrm{-C}_{3}^{}\mathrm{H}_{3}^{+} + \mathrm{e}^{-}  \rightarrow c\mathrm{-C}_{3}^{} \mathrm{H}_{2} + \mathrm{H}~.
\end{equation}
These reactions are less dependent on the temperature and can occur prior to CO freeze-out at higher temperatures and lower densities.
In Fig. \ref{fig:c_C3H2_1D_3D} we compare the radial abundance profiles of $c$-C$_3$H$_2$ and CH$_3$OH for the 1D model of \citet{2010MNRAS.402.1625K} with abundance profiles for the 3D core presented in this work. The lower panel shows an enhancement of the methanol abundance for the brown profile at $\sim$2500~au corresponding to the density and visual extinction enhancements seen in Fig. \ref{fig:av_1d_3d}. This confirms that regions with higher visual extinction and density also show a higher methanol abundance which may be offset from the center of the core. 
Figure \ref{fig:c_C3H2_1D_3D} shows a significant scatter among the different radial directions of the 3D model which results from the non-symmetrical structure of the core. The same effect is seen in Fig. \ref{fig:c_C3H2_kde}, where a Gaussian KDE is performed for the gas-phase $c$-C$_3$H$_2$ and CH$_3$OH abundances for all cells in the core as a function of the local extinction. The figure shows that many cells in a model with low visual extinction ($\log \mathrm{Av}$ < 0.4) have number densities of $c$-C$_3$H$_2$ of $\geq10^{-6}$~cm$^{-3}$. At higher visual extinction the number densities of $c$-C$_3$H$_2$ increases until around 10~mag where freeze-out starts to lower the absolute value. For CH$_3$OH the picture is different. Methanol is not formed at the lowest visual extinction, since CO does not fully freeze-out yet. From $\log \mathrm{Av}$ > 0.45 the number densities of CH$_3$OH reaches $\sim10^{-6}$~cm$^{-3}$ and at higher visual extinction methanol continues to reach higher number densities. Figure \ref{fig:c_C3H2_kde} also shows that c-C3H2 exhibits a stronger correlation with the local extinction, while a larger spread is seen in the CH$_3$OH panel at higher visual extinction. To understand this difference, we have to consider the different formation pathways of the two molecules. Methanol is formed through grain-surface reactions and desorbed into the gas-phase primarily through chemical desorption \citep{2016A&A...585A..24M,2017ApJ...842...33V}. $c$-C$_3$H$_2$ in the other hand is formed in the gas-phase through ion-neutral reactions. This difference in formation pathways leads to the different behavior at higher visual extinction: methanol formation is still active at higher visual extinction, with different formation rates depending on the local physical conditions, resulting in a continued release of methanol into the gas-phase through chemical desorption. For $c$-C$_3$H$_2$ the formation rate is primarily determined by the balance between the formation rate and the adsorption rate. Thus, the larger spread in gas-phase abundances for methanol stems from the stronger dependence on the local temperature and density for the grain-surface chemistry. 
Additional KDE figures for the cell density, temperature, and radius are shown in Appendix \ref{app:kde}.

\subsection{Comparing molecular emission maps}
\begin{figure*}[ht]
\resizebox{\hsize}{!}
        {\includegraphics{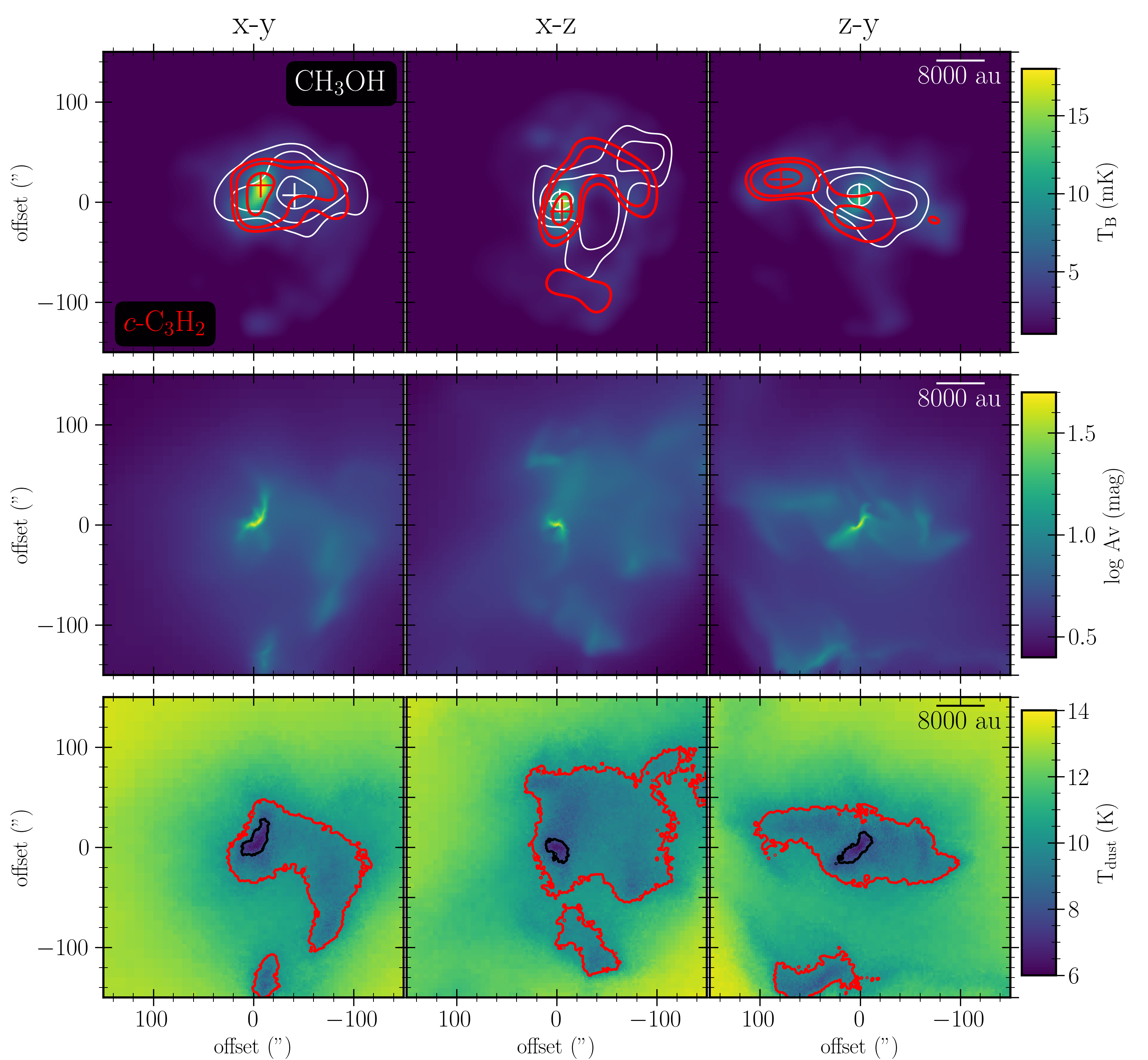}}
  \caption{Synthetic imaging and physical maps of the simulated core.
  \emph{Top)} Comparison between the location and spatial extent of the $c$-C$_3$H$_2$ 3$_{2,2}$--3$_{1,3}$  transition (red contours) and CH$_3$OH 2$_{1,2}$--1$_{1,1}$ transition (white contours) emission viewed from three different directions in the simulation. Contours show 60\%, 70\%, 80\%, and 90\% of the peak emission in the moment-zero maps for each transition. The continuum emission at 1.1~mm is shown by the color map.\newline 
  \emph{Middle)} The logarithm of the computed visual extinction in three planes of the core. A logarithmic scaling is applied to emphasize the variation across the core. A lower limit of A$_\mathrm{v} =$ 2~mag is applied.\newline 
  \emph{Bottom)} The computed dust temperature in three planes of the core. The black and red contours indicate the 8~K and 10~K temperature limits respectively. A lower limit of 6~K is applied.}
     \label{fig:3x3_LIME}
\end{figure*}

The top row of Fig. \ref{fig:3x3_LIME} shows a comparison between the morphology of the $c$-C$_3$H$_2$ 3$_{2,2}$--3$_{1,3}$ transition at 84.728~GHz and the CH$_3$OH 2$_{1,2}$--1$_{1,1}$ transition at 96.739~GHz, for the fiducial model. The contours show the structure of the integrated intensity (moment-zero) maps for each transition.
The three top panels of the figure show that the spatial location and extent of each molecular transition depends on the viewers location of the cloud core. The selected transitions are the same as those observed in \citet{2016A&A...592L..11S} toward L1544. In L1544, these molecules were found to trace different environments, notably related to the uneven illumination of the core. To test whether the molecular morphology in the present model shows a similar dependence on illumination, we present visual extinction and dust temperature maps of the core in the middle and bottom rows of Fig. \ref{fig:3x3_LIME}. Both panels show that the methanol emission is tracing the more shielded region of the core where the impact of photochemistry is reduced. In the top left panel of Fig. \ref{fig:3x3_LIME} (x-y plane), the methanol emission is shifted towards the right region of the dust continuum. This region also shows a drop in temperature and an increase in visual extinction, i.e., a higher shielding from the ISRF. The same trend is seen in the middle figure on the top row (x-y) where the methanol emission is again extended towards the right of the dust peak which is more shielded (i.e., lower dust temperature and higher visual extinction). For the third panel, methanol emission is mostly centered on the dust continuum, but still overlapping with the more shielded regions of the core.
For $c$-C$_3$H$_2$, the emission appears to trace a different gas component. In the x-y and x-z planes, the emission of $c$-C$_3$H$_2$ peaks close the dust peak while in the z-y plane the brightest emission is shifted far from the dust peak. The contours of the $c$-C$_3$H$_2$ and CH$_3$OH emission overlaps to a large extend in all planes, but the peak position varies considerably. 
  
To better quantify the observed patterns we computed the Pearson correlation coefficients $p$ between column density maps of $N$(H$_2$), $\hat{N}$(A$_\mathrm{v}$), $\hat{N}$(T$_\mathrm{gas}$), $N$(CH$_3$OH), and $N$($c$--C$_3$H$_2$). Here $N$ denotes the column density while $\hat{N}$ denotes the normalized column density. We used normalized column densities for visual extinction and gas temperature since the integrated column of these quantities are non-physical. 
The results are shown in Table \ref{tab:pearson}. First of all, the analysis shows correlation between the H$_2$ column density and integrated visual extinction ($\bar{p} = 0.73$ averaged across the three principal planes), as well as an anti-correlation between the integrated gas temperature and H$_2$ column density ($\bar{p} = -0.67$ averaged across the three principal planes). The analysis also shows correlations between $N$(H$_2$) and both $N$(CH$_3$OH) and $N$($c$--C$_3$H$_2$), of $\bar{p} = 0.62$ and $\bar{p} = 0.77$, respectively. Notably, a strong correlation of $\bar{p} = 0.92$ is found between the normalized column density of the visual extinction $\hat{N}$(A$_\mathrm{v}$) and $N$(CH$_3$OH), while a weaker correlation of $\bar{p} = 0.82$ is found between $\hat{N}$(A$_\mathrm{v}$) and $N$($c$--C$_3$H$_2$). Similarly, an anti-correlation is seen between the normalized integrated temperature and the two molecular species. While both molecules correlate with both the H$_2$ column density and the integrated visual extinction in the maps, the analysis shows that the carbon-chain molecule $c$--C$_3$H$_2$ is more strongly correlated to the H$_2$ column density than CH$_3$OH, while CH$_3$OH is more strongly correlated to the integrated visual extinction than $c$--C$_3$H$_2$.

\begin{table*}[]
\caption{Pearson correlation coefficients for the different column density maps. The coefficient is computed by comparing the different images pixel-by-pixel. The coefficient are listed for each of the three principal planes: x--y, x--z, z--y}\label{tab:pearson}
\centering
\begin{tabular}{lrrrrr}
\hline\hline
Quantity &  $N$(H$_2$) & $\hat{N}$(A$_\mathrm{v}$) & $\hat{N}$(T$_\mathrm{gas}$) & $N$(CH$_3$OH) & $N$($c$--C$_3$H$_2$) \\
\hline
$N$(H$_2$)        &   --    & 0.73, 0.73, 0.74  & -0.65, -0.67, -0.68  & 0.56, 0.63, 0.66  & 0.80, 0.75, 0.76  \\
$\hat{N}$(A$_\mathrm{v}$)        &  0.73, 0.73, 0.74     &  -- &  -0.99, -0.99, -0.99 & 0.93, 0.92, 0.91  & 0.84, 0.86, 0.77  \\
$\hat{N}$(T$_\mathrm{gas}$)       &   -0.65, -0.67, -0.68    & -0.99, -0.99, -0.99 &  -- &  -0.93, -0.91, -0.90 & -0.78, -0.81, -0.71  \\
$N$(CH$_3$OH)        &   0.56, 0.63, 0.66    &  0.93, 0.92, 0.91 & -0.93, -0.91, -0.90  &  -- &  0.68, 0.70, 0.61 \\
$N$($c$--C$_3$H$_2$)        &   0.80, 0.75, 0.76    &  0.84, 0.86, 0.77 &  -0.78, -0.81, -0.71 &  0.68, 0.70, 0.61 & --  \\
\hline
\end{tabular}
\end{table*}

\begin{figure*}[ht]
\resizebox{\hsize}{!}
        {\includegraphics{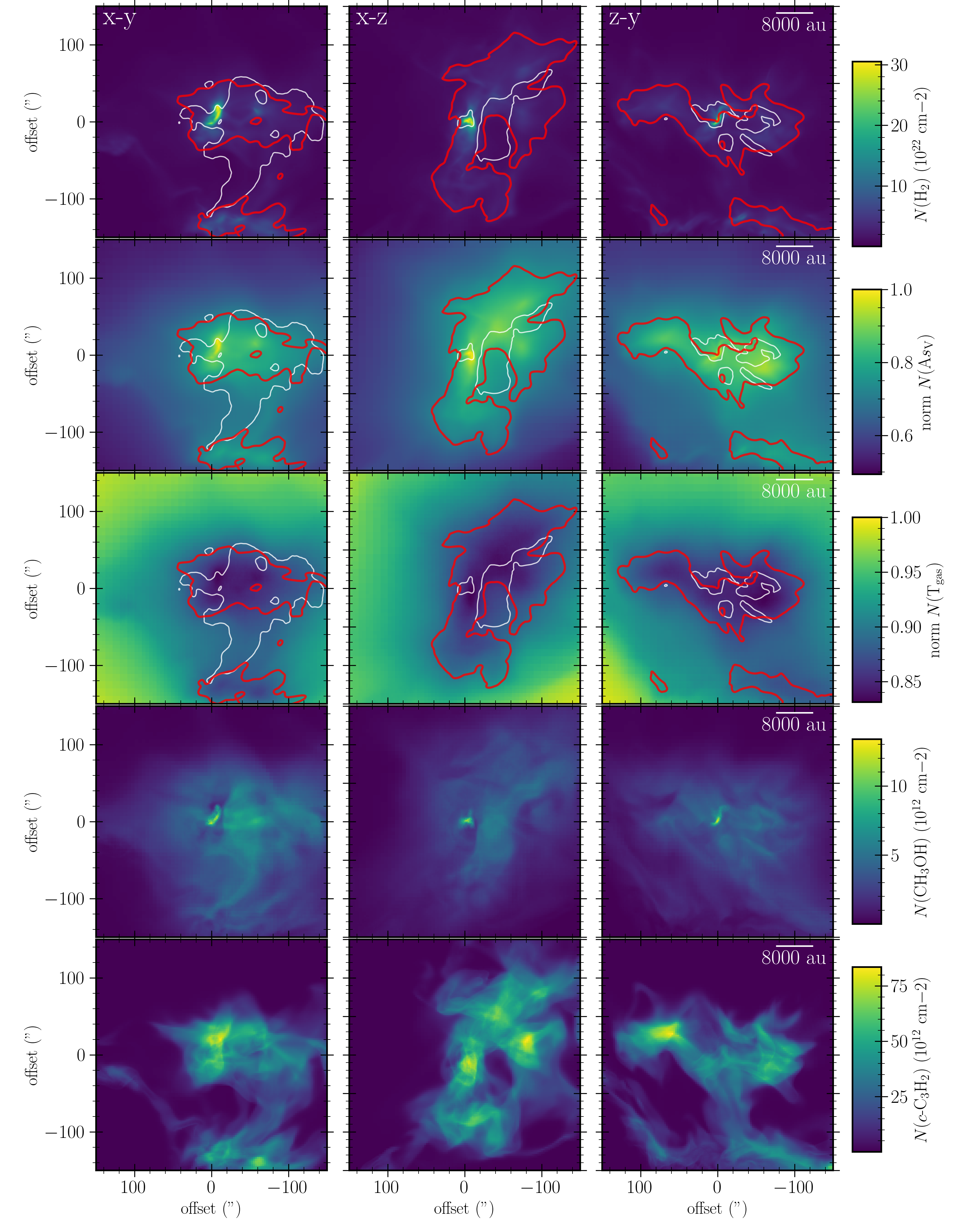}}
  \caption{Column density maps for $N$(H$_2$), $\hat{N}$(A$_\mathrm{v}$), $\hat{N}$(T$_\mathrm{gas}$), $N$(CH$_3$OH), and $N$($c$--C$_3$H$_2$). All maps are for the fiducial model. In the top 3 rows the contours indicate regions with column densities above 30\% of the peak column density for CH$_3$OH (white) and $c$--C$_3$H$_2$ (red).}
     \label{fig:5x3}
\end{figure*}




\subsection{Comparing different chemical models}
\begin{figure*}[ht]
\resizebox{\hsize}{!}
        {\includegraphics{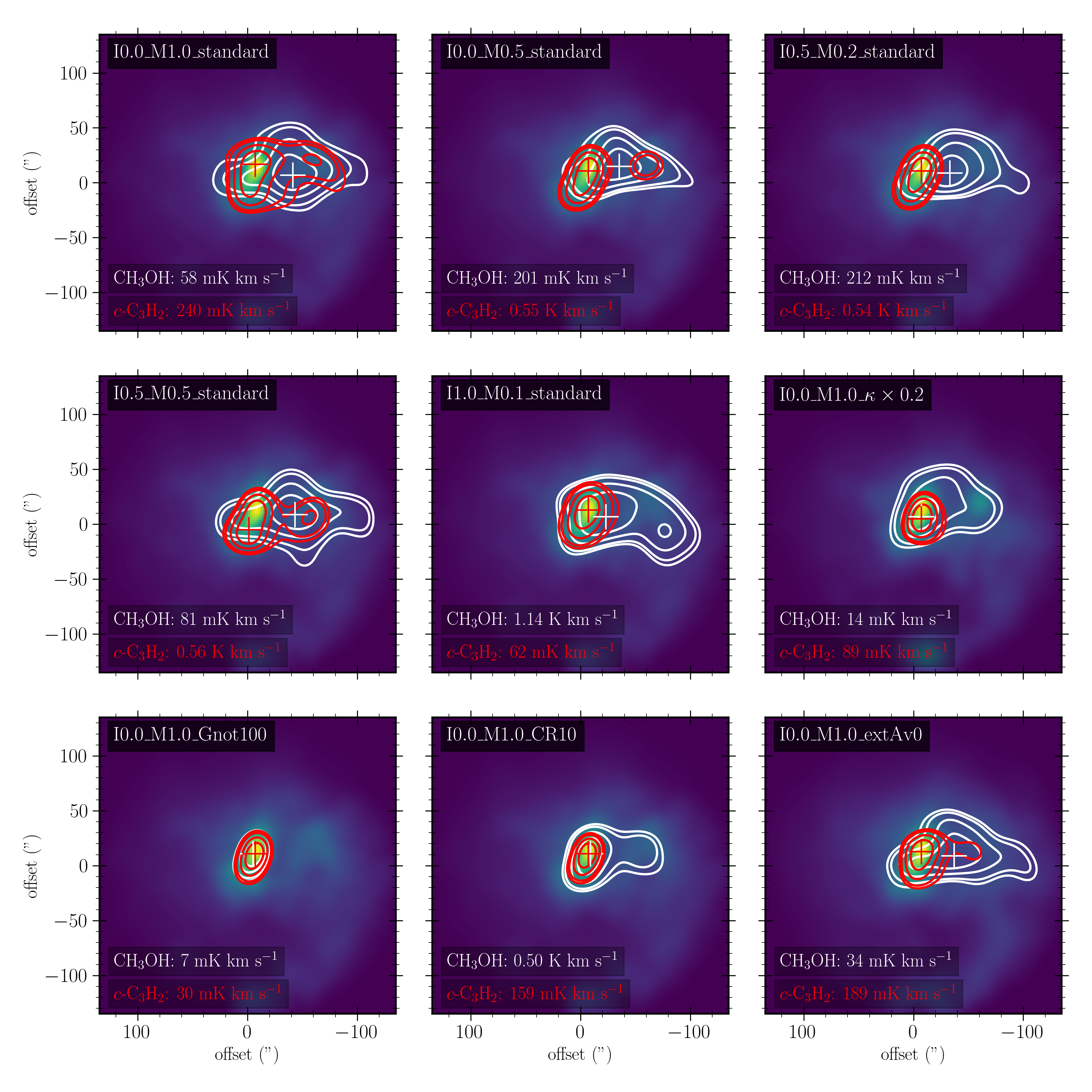}}
  \caption{Comparison between the $c$-C$_3$H$_2$ 3$_{2,2}$--3$_{1,3}$ transition (red contours) and CH$_3$OH 2$_{1,2}$--1$_{1,1}$ (white contours) integrated emission for each of the 9 chemical models listed in Table \ref{tab:chemical_models}. The figure shows the x-y plane seen in the first column of Fig. \ref{fig:3x3_LIME}.}
     \label{fig:model_comparison}
\end{figure*}

The exact chemical age and initial conditions for the chemistry in L1544 is uncertain and we have therefore tested a number of different initial conditions for the chemical model, as introduced in Table \ref{tab:chemical_models}. In Fig. \ref{fig:model_comparison}, we show the $c$-C$_3$H$_2$ and CH$_3$OH emission in contours of red and white, respectively, for each of the models presented in Table \ref{tab:chemical_models}. The integrated intensity of the line is also included for each molecule. 
The chemical morphology for these molecular emission lines show a similar morphology for models $\mathrm{I0.0\_M1.0\_standard}$, $\mathrm{I0.5\_M0.2\_standard}$, $\mathrm{I0.5\_M0.5\_standard}$, and $\mathrm{I1.0\_M0.1\_standard}$, with methanol peaking toward the right of the core and $c$-C$_3$H$_2$ peaking close to the dust peak. Models $\mathrm{I0.0\_M1.0\_}\kappa\times0.2$ and $\mathrm{I0.0\_M1.0\_CR10}$ also shows a similar offset in the methanol contours, but the emission peaks on the dust peak and not offset to the right. The remaining models show a different morphology. Model $\mathrm{I0.0\_M1.0\_Gnot100}$ show very faint CH$_3$OH emission which peaks on the dust peak. Models $\mathrm{I0.0\_M0.5\_standard}$ and $\mathrm{I0.0\_M1.0\_extAv0}$ also show faint CH$_3$OH emission and a brighter $c$-C$_3$H$_2$ with a larger extent. Models $\mathrm{I0.0\_M0.5\_standard}$, $\mathrm{I0.0\_M1.0\_}\kappa\times0.2$, $\mathrm{I0.0\_M1.0\_Gnot100}$, and $\mathrm{I0.0\_M1.0\_extAv0}$ are all inconsistent with observations of L1544 due to the weak CH$_3$OH emission or the morphology of the emission. The stronger line strengths of $c$-C$_3$H$_2$ in models $\mathrm{I0.0\_M0.5\_standard}$, $\mathrm{I0.5\_M0.2\_standard}$, and $\mathrm{I0.5\_M0.5\_standard}$ is consistent with carbon-chain molecules forming earlier than CH$_3$OH since these models are more chemically young. The behavior of models $\mathrm{I0.0\_M1.0\_}\kappa\times0.2$ and $\mathrm{I0.0\_M1.0\_extAv0}$ are comparable as the effect of the changed opacity on the temperature structure is comparable to that of the lower visual extinction in model $\mathrm{I0.0\_M1.0\_extAv0}$, although the impact is strongest in $\mathrm{I0.0\_M1.0\_}\kappa\times0.2$ where methanol formation is weaker. 
Models with very weak CH$_3$OH emission are not consistent with observations of L1544 which shows relatively bright CH$_3$OH emission (See also Table \ref{tab:fit_spectra} and the section below).

In Appendix \ref{app:3} we have included similar figures to Fig. \ref{fig:model_comparison} for all three planes.

\subsection{Comparison of spectra for CH$_3$OH and $c$-C$_3$H$_2$}
\begin{figure}[ht]
\resizebox{\hsize}{!}
        {\includegraphics{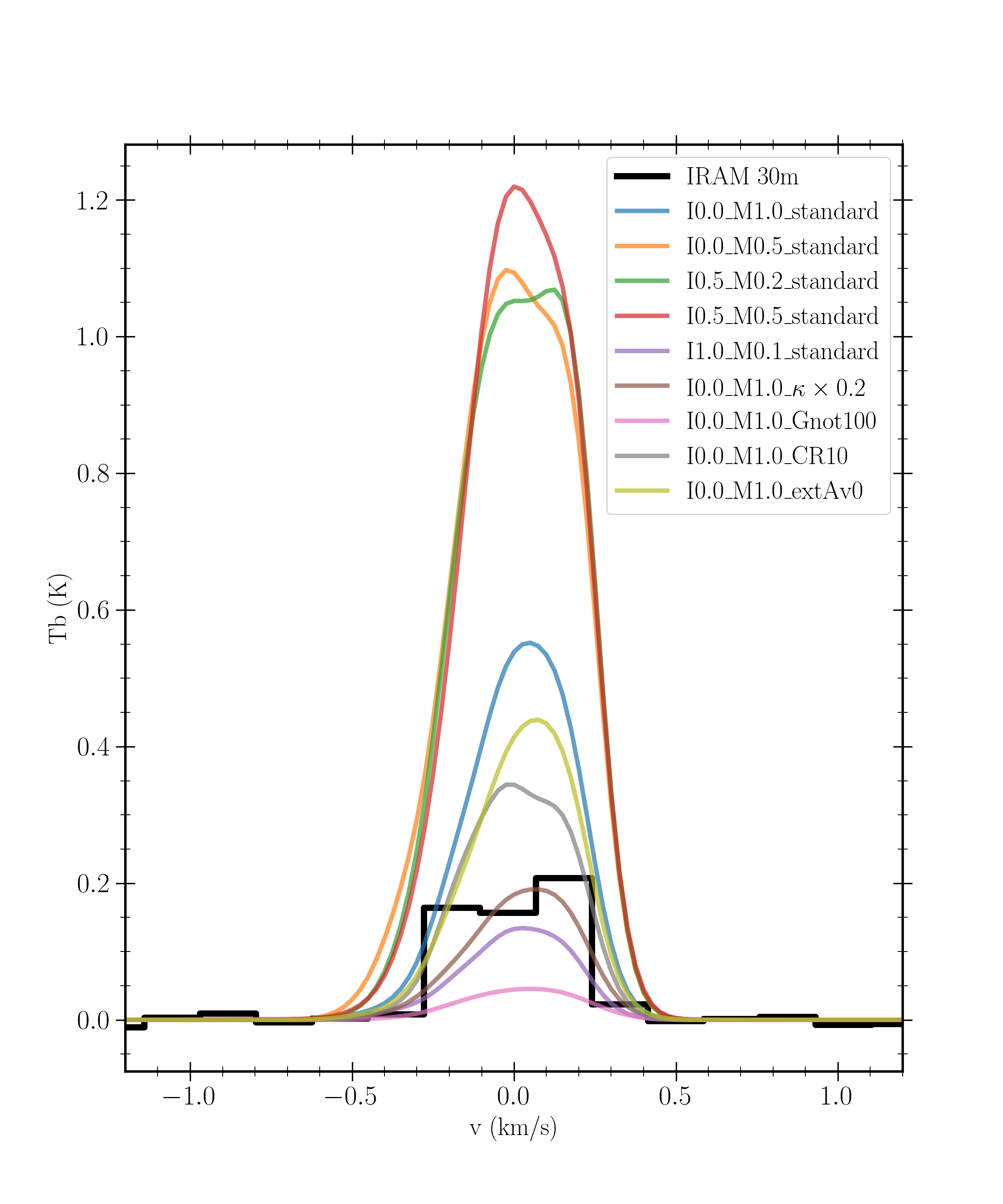}}
  \caption{Comparison between the $c$-C$_3$H$_2$ 3$_{2,2}$--3$_{1,3}$ spectrum for each of the 9 chemical models listed in Table \ref{tab:chemical_models} and the observed spectrum from IRAM 30m observations \citep{2016A&A...592L..11S}.}
     \label{fig:C3H2_spectra}
\end{figure}

\begin{figure}[ht]
\resizebox{\hsize}{!}
        {\includegraphics{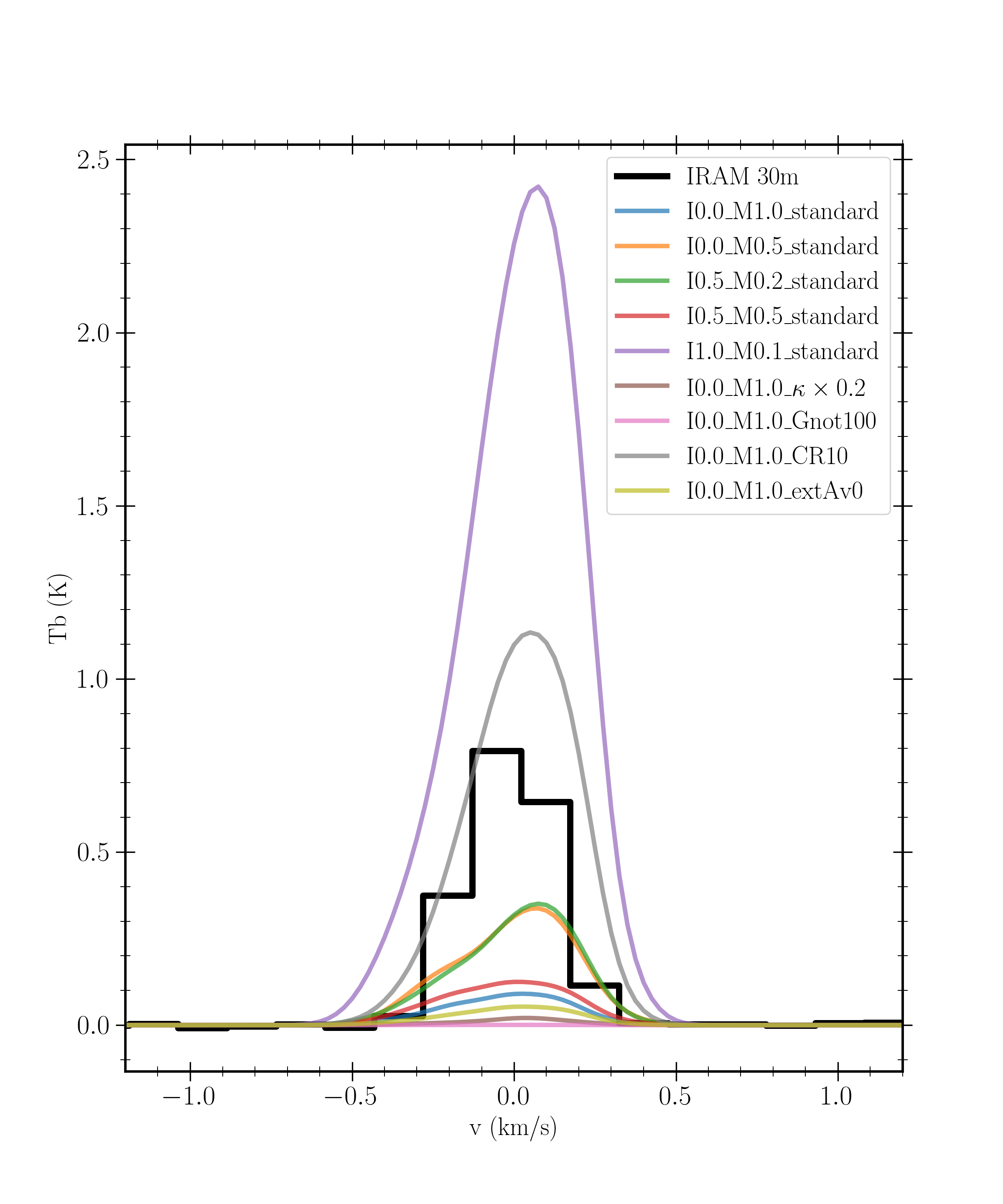}}
  \caption{Comparison between the CH$_3$OH 2$_{1,2}$--1$_{1,1}$ spectrum for each of the 9 chemical models listed in Table \ref{tab:chemical_models} and the observed spectrum from IRAM 30m observations \citep{2016A&A...592L..11S}.}
     \label{fig:CH3OH_spectra}
\end{figure}

Figure \ref{fig:C3H2_spectra} presents a comparison between the $c$-C$_3$H$_2$ 3$_{2,2}$--3$_{1,3}$ spectrum towards the dust peak of L1544 and towards the dust peak of the simulated L1544 analog for the nine different chemical models and Figure \ref{fig:CH3OH_spectra} is similar for the CH$_3$OH 2$_{1,2}$--1$_{1,1}$ transition. An overview of the modeled spectra including fitted Gaussian parameters is included in Table \ref{tab:fit_spectra} for both transitions. We fitted Gaussian profiles to the modeled lines although some of the spectra deviate from a pure Gaussian profile due to asymmetries related to self-absorption in the infalling core. This is done because the spectral resolution of the observed spectra toward L1544 is insufficient to resolve these features.

The models show a broad range of intensities for the targeted lines, depending both on the choice of chemical age, ISRF ($G_0$), and the cosmic-ray ionization rate. For the fiducial model the peak brightness for $c$-C$_3$H$_2$ is $\sim$3 times higher than what is observed towards L1544, while the integrated line intensity is higher by a factor of $\sim$2.5. The width of the modeled lines are generally consistent to the observed line (0.47 km/s compared to 0.38-0.44 km/s), indicating the the kinematics of the modeled core is comparable to L1544 in the region traced by this $c$-C$_3$H$_2$ transition. However, the limited spectral resolution of the observed line prohibits more detailed comparison.
In terms of the integrated intensities, models $\mathrm{I0.0\_M1.0\_standard}$, $\mathrm{I1.0\_M0.1\_standard}$, $\mathrm{I0.0\_M1.0\_}\kappa\times0.2$, $\mathrm{I0.0\_M1.0\_CR10}$, and $\mathrm{I0.0\_M1.0\_extAv0}$ provide emission lines which agree to less than a factor of three with the observed line. 
For CH$_3$OH, the fiducial model has a too low peak brightness by a factor of $\sim$8. Models $\mathrm{I0.0\_M0.5\_standard}$, $\mathrm{I0.5\_M0.2\_standard}$ and $\mathrm{I0.0\_M1.0\_CR10}$ provide the closest match to the observed peak brightness (0.33~K and 1.16~K respectively compared to the observed 0.84~K). Model $\mathrm{I0.0\_M1.0\_Gnot100}$ has a too low methanol abundances to provide a decent Gaussian fit. Generally, the majority of the models show weaker methanol emission than what is observed. This could be due to missing (non-thermal) desorption pathways in the shielded regions of the core.

To achieve a better match, the duration of the chemical model could be fine-tuned to fit each specific molecule, but this is beyond the scope of this work where we focus on the chemical structure of the core. Finally, we note that differences in the physical structure of the modeled core and L1544 could limit the degree to which the observed spectra can be reproduced.

\begin{table*}
\caption{Overview of the modeled spectra for $c$-C$_3$H$_2$ and CH$_3$OH shown in Figures \ref{fig:C3H2_spectra} and \ref{fig:CH3OH_spectra}. The columns list the peak brightness, the FWHM, and the area of the emission lines. All spectra are extracted toward the dust peak and fitted  with Gaussian profiles. Parameters for a Gaussian fit to the observed spectra toward L1544 is included in the top row.}\label{tab:fit_spectra}
\centering
\begin{tabular}{llllclll}
\hline\hline
& & $c$-C$_3$H$_2$ & & & & CH$_3$OH  & \\ 
Model & $T_\mathrm{peak}$ (K) & $\mathrm{d}v$ (km/s) & $W$ (K km${}^{-1}$) & & $T_\mathrm{peak}$ (K) & $\mathrm{d}v$ (km/s) & $W$ (K km${}^{-1}$) \\ 
 \hline
L1544   & 0.21                  & 0.47                      & 0.10       & & 0.84 & 0.33 & 0.30     \\
$\mathrm{I0.0\_M1.0\_standard}$   & 0.57                  & 0.39                      & 0.23        & & 0.09 & 0.45 & 0.04     \\
$\mathrm{I0.0\_M0.5\_standard}$    & 1.16                  & 0.44                      & 0.55     & & 0.33 & 0.44 & 0.15 \\
$\mathrm{I0.5\_M0.2\_standard}$   & 1.16                  & 0.44                      & 0.55   & & 0.33 & 0.42 & 0.15 \\
$\mathrm{I0.5\_M0.5\_standard}$   & 1.26                  & 0.41                      & 0.56      & & 0.13 & 0.48 & 0.07    \\
$\mathrm{I1.0\_M0.1\_standard}$   & 0.14                  & 0.39                      & 0.06    & & 2.4 & 0.41 & 1.06 \\
$\mathrm{I0.0\_M1.0\_}\kappa\times0.2$   & 0.20                  & 0.40                      & 0.08       & & 0.02 & 0.38 & 0.01   \\
$\mathrm{I0.0\_M1.0\_Gnot100}$   & 0.05                  & 0.43                      & 0.02      & & -- & -- & --  \\
$\mathrm{I0.0\_M1.0\_CR10}$   & 0.36                  & 0.41                      & 0.16     & & 1.16 & 0.40 & 0.49  \\
$\mathrm{I0.0\_M1.0\_extAv0}$   & 0.45                  & 0.38                      & 0.18     & & 0.05 & 0.45 & 0.03  \\
\hline
\end{tabular}
\end{table*}

\section{Discussion}  \label{sec:discussion}
With the development of this 3D physico-chemical model we aim to understand the origin of the observed chemical differentiation in L1544 and other pre-stellar cores. Most notably, an offset between the dust peak and the methanol peak has been reported in several starless and pre-stellar cores \citep[e.g., ][]{2020A&A...643A..60S, 2020ApJ...895..101H}. The observed chemical distribution could result from uneven illumination of the cores which are not considered in the one-dimensional models regularly used to study these cores. Uneven illumination can arise from a number of different situations. In \citet{2020A&A...643A..60S}, an offset between methanol and cyclopropenylidene is reported for five out of six cores, where the sixth core showing no offset is the isolated Bok globule B68, thus uniformly illuminated. The remaining cores show an offset that is due to some form of uneven illumination of the cores. The cause of the uneven illumination varies from intrinsic variations in the density distribution to external illumination from nearby massive B-type stars. As such, uneven illumination seem to be almost ubiquitous in cores located in star-forming regions, with several different factors contributing to the effect. Indeed, \citet{2022A&A...657A..10S} studied the methanol distribution around 12 cores in the Taurus molecular cloud and found two different dichotomies related to the environment of the core. In one group, the N(CH$_3$OH)/N(C$^{18}$O) ratio peaks at the dust peak, while for the other group, the peak is offset by $\sim$10,000~au. The first group is identified as sources located in regions with higher rates of star formation which can influence the formation of methanol from CO in the outer layers of the core. Nearby protostars can suppress the formation of methanol through higher cosmic-ray fluxes from particles launched in outflow and jets \citep{2016A&A...590A...8P, 2021ApJ...915...43F} or through higher rates of photodesorption, limiting CO freeze-out. Furthermore, the presence of outflows from the surrounding protostars will create a clumpier medium where the ISRF may penetrate more efficiently. The second group is located in regions with lower rates of star formation and as such these effects are diminished.
In the case of L1544, the uneven illumination of the core is inferred from the sharp drop in the N(H$_2$) column density on one side of the core \citep{2016A&A...592L..11S}. In the model presented here, we apply a uniform ISRF around the computational domain. Thus, the uneven illumination in the 3D model arises from the physical structure of the core within the computational domain (r $\leq$ 50,000~au), i.e., at the core-cloud boundary and within the core itself. In the upper right panel in Fig. \ref{fig:av_1d_3d} a notable difference in the visual extinction is seen on scales ranging from $\sim$1000~au to the edge of the domain at 50,000~au (e.g., from $\sim$6~mag to 30~mag at 2,000~au). Variations are also evident in the density profiles and to a lesser extent in the temperature profiles. These variations arise from the clumpy structure of the core and envelope and are the cause of the observed chemical dichotomy of the core shown in Fig. \ref{fig:3x3_LIME}.


\subsection{Comparison with the observed chemical morphology of $c$-C$_3$H$_2$ and CH$_3$OH in L1544}

The observed chemical morphology of $c$-C$_3$H$_2$ and CH$_3$OH in L1544 is shown in Fig. \ref{fig:observations}. The methanol emission peaks toward the north-west of the dust peak where the column density is higher than toward the south-east. This is consistent with the methanol emission in our model, which peaks on the right side of the pre-stellar core in the x-y and x-z planes where the visual extinction is higher and the dust temperature lower, i.e., the more shielded side of the core. Furthermore, this correlation is also evident from the Pearson correlation coefficients, as mentioned in Section \ref{sec:results}. Meanwhile, the $c$-C$_3$H$_2$ contours in the emission map of L1544 overlaps with the continuum, but peaks toward the southern part of the core. This is also the case for our model where the contours of $c$-C$_3$H$_2$ are centered around the dust peak. Our model predicts that $c$-C$_3$H$_2$ is more abundant in the chemically younger models similar to other carbon-chain molecules \citep[e.g.,][]{2013ChRv..113.8981S}.

\citet{2016A&A...592L..11S} suggested that the observed offset between CH$_3$OH, $c$-C$_3$H$_2$, and the dust peak of L1544 was the result of an uneven irradiation of the core on different sides. The synthetic observations presented here support this explanation since the CH$_3$OH emission peaks on the more shielded side of our L1544 analog. This is a consequence of the more efficient freeze-out of CO on this side of the core, which leads to a higher production of CH$_3$OH through grain-surface chemistry. The desorption of methanol in the dense and cold core studied here is driven mainly by chemical desorption with insignificant contributions from cosmic-ray induced desorption and photodesorption. Consequently, the modeled gas-phase abundances depend sensitively on the assumed efficiency of the chemical desorption of methanol. In this work we followed the description of \citet{2007A&A...467.1103G} for all reactions, however the efficiency can vary which could impact the gas-phase abundances considerable \citep[e.g.,][]{2016A&A...585A..24M}. This would however not impact the morphology of the methanol emission which is the main focus of this work.
These results confirm that the core structure and external environment has a notable impact on the methanol distribution in the core by influencing the strength of the local irradiation. Furthermore, the model predicts that the brightness of CH$_3$OH depends strongly on the choice of chemical age and the UV irradiation. Models with lower shielding or stronger ISRF ($\mathrm{I0.0\_M1.0\_}\kappa\times0.2$, $\mathrm{I0.0\_M1.0\_Gnot100}$, and $\mathrm{I0.0\_M1.0\_extAv0}$) show very faint CH$_3$OH emission at the dust peak. 

For $c$-C$_3$H$_2$, the location of the peak molecular emission in L1544 is closer to the high column density structure, peaking on the southern side of the core. In our model, the targeted $c$-C$_3$H$_2$ transition also peaks close to the dust peak in the x-y and x-z planes (top left and middle figures, Fig. \ref{fig:3x3_LIME}). However, the $c$-C$_3$H$_2$ emission is shifted $\sim$8000~au from the dust peak in the z-y plane. The variation between the different planes highlights that caution is needed when interpreting observations. Combined emission maps and multi-line analysis can help to constrain the emission region \citep[e.g.,][]{2022A&A...665A.131L}. Comparing the synthetic observations in Fig. \ref{fig:3x3_LIME} with the column density maps in Fig. \ref{fig:5x3}, it is evident that $c$-C$_3$H$_2$ is more extended than CH$_3$OH, but the $c$-C$_3$H$_2$ 3$_{2,2}$--3$_{1,3}$ transition shown in Fig. \ref{fig:3x3_LIME} predominantly traces the inner part of the core due to excitation effects (critical density > 10$^{5}$~cm${}^{-3}$.
When comparing the different chemical models in Fig. \ref{fig:model_comparison}, only minimal offsets ($<$ 20${}^{''}$, $<$ 4000~au) between the $c$-C$_3$H$_2$ contours are present in the different chemical models, indicating that the distribution of $c$-C$_3$H$_2$ in the core does not depend strongly on the choice of chemical initial conditions and external fields.

\begin{figure}[ht]
\resizebox{\hsize}{!}
        {\includegraphics{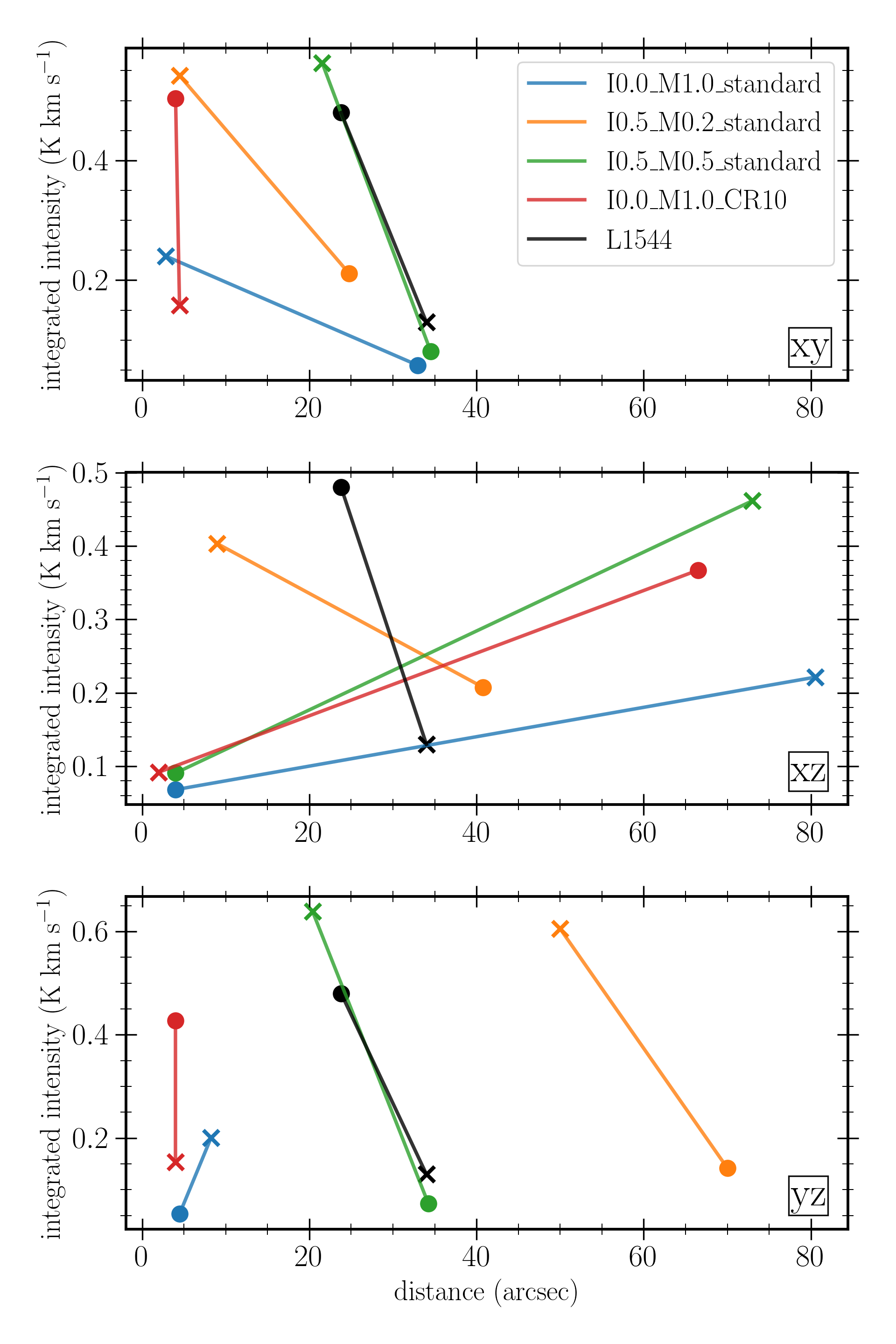}}
  \caption{Comparison of the molecular offsets and brightness for $c$-C$_3$H$_2$ 3$_{2,2}$--3$_{1,3}$ transition (crosses) and CH$_3$OH 2$_{1,2}$--1$_{1,1}$ (circles) for selected chemical models and L1544. The x-axis shows the distance between the dust peak and the molecular emission peak in arcseconds and the y-axis shows the integrated intensity for the molecules.}
     \label{fig:distance_plot}
\end{figure}

To provide a quantitative overview, Figure \ref{fig:distance_plot} shows a comparison between the molecular offsets and integrated intensities for selected models and the observed values for L1544 from \citet{2016A&A...592L..11S}. Focusing on the top panel with the x-y plane, the offsets for methanol are comparable to the observed value for L1544 (circles), except for model $\mathrm{I0.0\_M1.0\_CR10}$. However, this is not the case in the remaining planes where a substantial scatter is visible. For $c$-C$_3$H$_2$ (crosses), the distance from the dust peak is larger in L1544 then the model in most models and planes. Outliers are seen in plane z-y where the $c$-C$_3$H$_2$ emission shows a large offset of more than 50${}^{''}$. Model I$0.0\_$M$1.0\_$CR$10$ provide the best match to the observed integrated intensities of both molecules, however it does not provide a good match for the observed chemical distribution, regardless of the imaged plane. This is also seen in Fig. \ref{fig:3x3_LIME}. We note that while the present model does not reproduce the offset of the peak $c$-C$_3$H$_2$ position observed in L1544 very well, the extent of the contours show a similar trend for both models and observations, with the contours of the $c$-C$_3$H$_2$ emission centered on the dust continuum.

\subsection{Caveats of the model}
While the present model includes several improvements over static or dynamic 1D models of L1544 and pre-stellar cores in general, a number of caveats still remain. 
In the present model, we do not account for the dynamical evolution of the core, i.e., the gradual increase of density and decrease in temperature as the molecular cloud cools and contracts while forming the core. Following the dynamical evolution of the core when computing the chemical abundances may impact the abundance of certain species which is not fully captured in the present setup. However, if the chemical evolution is followed over time, a three-phase chemical model (i.e., separation of ice surface and ice mantle chemistry) may also be needed since two-phase models cannot record the chemical evolution in different ice layers. Such a change would further enhance the computational demands of the model.
The omission of a dynamical core evolution means that we cannot predict the chemical age of the core and instead assumed several different values. Future work can address this by calculating the chemical evolution along tracer particles which trace the gas evolution during the evolution of the pre-stellar core and allow a more realistic comparison between the modeled and observed chemical abundances and morphologies \citep[e.g.,][]{2014MNRAS.445..913D, 2017A&A...599A..40F, 2021A&A...649A..66J, 2021MNRAS.505.3442F}.

The present model does not include thermochemistry, i.e., the heating and cooling of molecular gas through absorption and emission. Furthermore, the physical evolution as computed using {\sc ramses} assumes an isothermal gas with no feedback from the microphysics. These simplifications may impact the dynamical evolution of the model and to some extent the structure of the core. Instead, the model presented here includes the large-scale turbulence of the molecular cloud. 
This approach is supported by the success of the model in reproducing the star formation rate, the initial mass function, and the core mass function \citep{2018ApJ...854...35H, 2021MNRAS.504.1219P}.

The present chemical model is limited to a two-phase gas-grain model as opposed to the more computationally expensive three-phase models \citep[e.g.,][]{1993MNRAS.263..589H, 2017ApJ...842...33V, 2017A&A...599A..40F}. This choice impacts the abundances of molecules formed through grain-surface chemistry, such as methanol. In the case of our fiducial model where the chemistry was evolved for 1 Myr, the difference in abundances between the two-phase and three-phase model is low for the current network, however the differences for different chemical ages can be significantly larger. However, as this work does not aim to model the exact abundances of CH$_3$OH and $c$-C$_3$H$_2$ the conclusions in this paper would not be impacted be a change to three-phase gas-grain model.

Another limitation of the model is the choice of a fixed cosmic-ray ionization rate throughout the core. This is a common assumption in astrochemical models \citep[e.g.,][]{2015A&A...578A..55S, 2020ApJ...897..110A}. However, cosmic-rays are expected to attenuate at higher optical depths in pre-stellar cores such as L1544 \citep[e.g.,][]{2018A&A...619A.144P, 2021A&A...656A.109R}. This effect could reduce the desorption of CH$_3$OH in the central part of the core where the cosmic-ray induced desorption is a significant factor in the model. We nonetheless choose to use a constant cosmic-ray flux in this work to limit the complexity of the model and because the degree of attenuation of cosmic rays in cloud cores remains uncertain \citep{2019ApJ...879...14S}.

A 3D model of L1544 was recently presented by \citet{2023MNRAS.521.2833G} to address the chemical dichotomy of L1544. The aims of that work is similar to the work presented here but the methods differ. In \citet{2023MNRAS.521.2833G}, the 3D structure is inferred from the 2D density structure in the plane of the sky using the Aviator code \citep{2020A&A...633A.132H}. The model assumes an isothermal core and the extinction was estimated from the integrated H$_2$ column density instead of a direct computation using radiative transfer modeling. As such the two models are complementary and show two different approaches to the 3D problem. Nonetheless, both models find that the irregular structure of the 3D cores can induce the chemical dichotomy observed in L1544.

\section{Summary}  \label{sec:5}
This work presents a 3D MHD physico-chemical model of a pre-stellar core embedded in a realistic molecular cloud environment. The model self-consistently computes the temperature and visual extinction in the core which allows us to study the impact of the local cloud environment and core structure on the chemical morphology in the core. The model is compared with observations of the pre-stellar core L1544 in order to explain the chemical structure of L1544.
The main results of the paper are as follows.

\begin{itemize}
    \item The transition from a spherically symmetric 1D model to a more realistic 3D model embedded in a molecular cloud leads to a chemical differentiation between CH$_3$OH and $c$-C$_3$H$_2$ which can be explained by the uneven illumination within the pre-stellar core and envelope. This is consistent with the observed dichotomy in L1544 and supports the explanation provided in \citet{2016A&A...592L..11S}. Uneven illumination occurs in this model as a direct consequence of the realistic environment of the molecular cloud simulation which determines the core and envelope structure and asymmetries which provoke the chemical structure.
    
    \item The observed chemical differentiation between CH$_3$OH and $c$-C$_3$H$_2$ is present for a broad range of different initial conditions for the chemical model, when standard values for the ISRF ($G_0 = 1$) and cosmic-ray flux ($\zeta_\mathrm{CR} = 3\times10^{-17}$ s$^{-1}$) are used. However, spectral comparison between the modeled core and L1544 demonstrates that the model presented here needs fine-tuning to better reproduce the brightness and line profile of both molecules. 

    \item The observed brightness and dichotomy of CH$_3$OH and $c$-C$_3$H$_2$ emission in L1544 cannot be reproduced by models with an enhanced ISRF of $G_0 = 100$ or an increased cosmic-ray ionization rate of 3$\times10^{-16} $s$^{-1}$.
    
    \item Comparison between the 1D and 3D models shows that the inner core structure (r $< 1000$~au) shows a similar structure and temperature profile. On larger scales the 3D model shows considerable scatter between radial profiles. We show in this work that these non-spherical features lead to the chemical dichotomy of the core as observed in L1544 and other cores \citep{2016A&A...591L...1S, 2020A&A...643A..60S}. It is therefore necessary to consider non-spherical features when interpreting molecular emission maps. Nonetheless, the mean profile averaged over different directions remains close to the 1D model.
    
\end{itemize}

This work demonstrates how the local cloud structure can directly impact the chemical structure of pre-stellar cores and highlights the importance of realistic 3D models when studying the core-envelope transition of young cores. This conclusion is further supported by the observations of CH$_3$OH and $c$-C$_3$H$_2$ presented in \citet{2016A&A...592L..11S} and \citet{2020A&A...643A..60S}. Through chemical inheritance, the imprinted chemical segregation can impact the system at later stages of the star- and planet-formation process, particularly at the onset of planet formation.
Future work will extent this study to include the dynamics of the gas and broaden the molecular tracers studied. With this model, we can also study the accretion process onto the core and analyze the formation and evolution of streamers which may play an important role in the physical and chemical evolution of protostellar envelopes and protoplanetary disks \citep[e.g.,][]{2020NatAs...4.1158P, 2022A&A...667A..12V, 2023A&A...670L...8G, 2023K}.

\begin{acknowledgements}
We wish to thank the anonymous referee for their thoughtful comments, which
improved the paper.
S.S.J. and S.S. wish to thank the Max Planck Society for the Max
Planck Research Group funding. All others authors affiliated to the MPE wish to
thank the Max Planck Society for financial support.
The research leading to these results has received funding from the Independent Research Fund Denmark through grant No. DFF 8021-00350B (TH). The astrophysics HPC facility at the University of Copenhagen, supported by research grants from the Carlsberg, Novo, and Villum foundations, was used for carrying out the simulations, the analysis, and long-term storage of the results.
\end{acknowledgements}

%
%

\bibliographystyle{aa}
\bibliography{b.bib}

\begin{appendix} 
\FloatBarrier
\section{Comparing interpolation and direct chemical modeling on 1D model of L1544.}\label{app:0}
\begin{figure}[ht]
 \resizebox{\hsize}{!}{\includegraphics{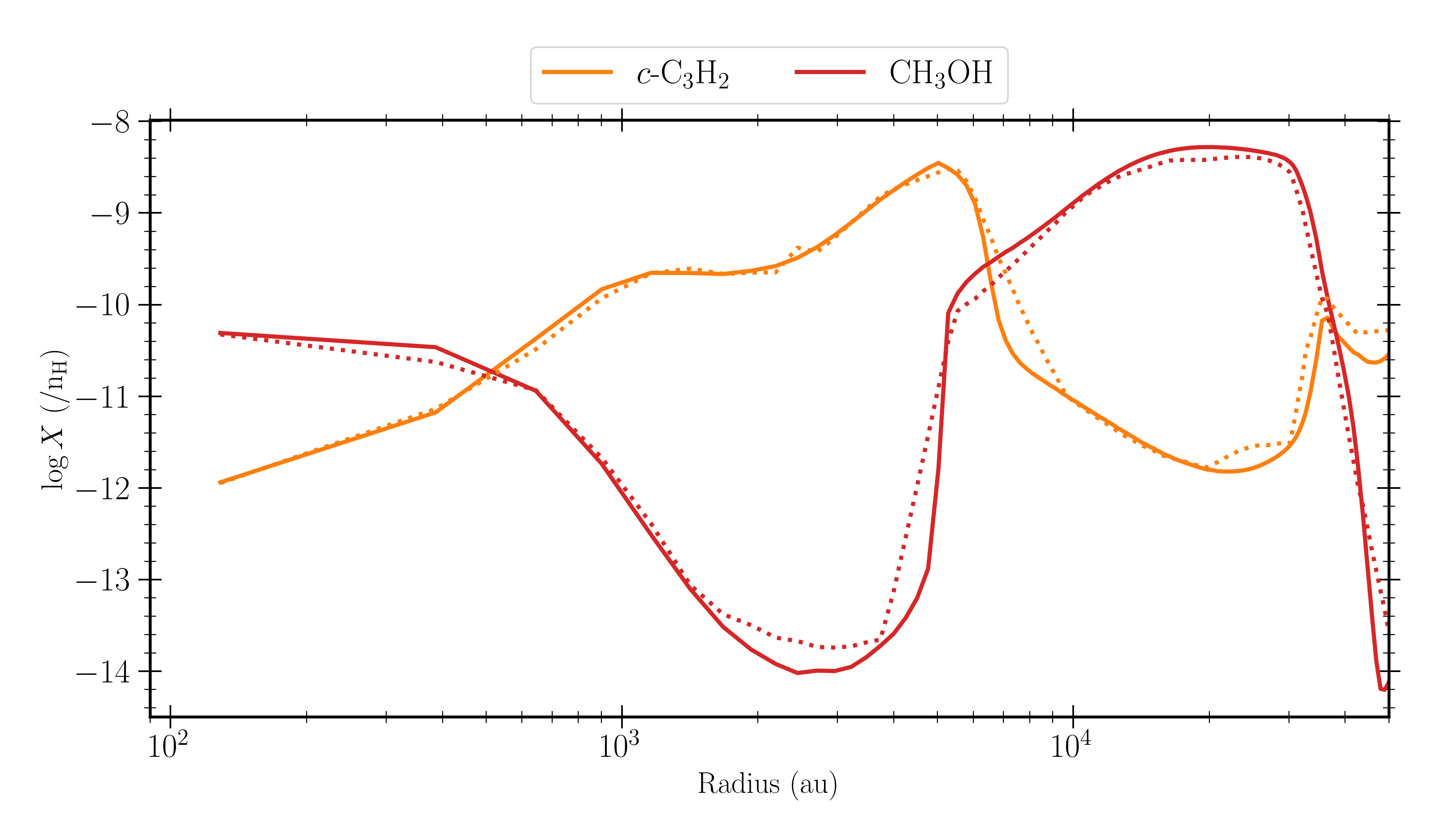}}
  \caption{Comparison between the radial chemical profiles when computing the chemical abundances directly on the 1D model grid (solid lines) and when using the interpolation module (dotted lines) for CH$_3$OH and $c$-C$_3$H$_2$. } 
     \label{fig:spectrum_HCN}
\end{figure}
\FloatBarrier

\section{Surface reactions with custom barriers}\label{app:reactions}
\FloatBarrier
\begin{table}
\caption{Reactions with custom barrier widths. Note that deuterated variants of the reactions have been excluded from the list, but share the same barrier widths.}
\centering
\begin{tabular}{lll}
\hline\hline
Reaction & Barrier width (\AA) & Reference \\\hline
O + CO $\rightarrow$ CO$_2$ & 1.25 & a,b  \\
H + CH$_4$ $\rightarrow$ CH$_3$ + H$_2$ & 2.17 & a,b \\
H + CO $\rightarrow$ HCO & 2.0 & b \\
H + H$_2$CO $\rightarrow$ CH$_2$OH & 2.0 & b \\
H + H$_2$CO $\rightarrow$ CH$_3$O & 2.0 & b \\
H + CH$_3$OH $\rightarrow$ CH$_2$OH + H$_2$ & 2.0 & b \\
H + CH$_3$OH $\rightarrow$ CH$_3$O + H$_2$ & 2.0 & b \\
\hline
\end{tabular}
\label{tab:reactions}
\tablebib{
(a) \citet{garrod2011}; (b) \citet{garrod2013}; (c) \citet{aikawa2012}}
\end{table}
\FloatBarrier

\section{Radial profiles for the core.}\label{app:2}
\FloatBarrier
\begin{figure*}[ht]
\resizebox{\hsize}{!}
        {\includegraphics{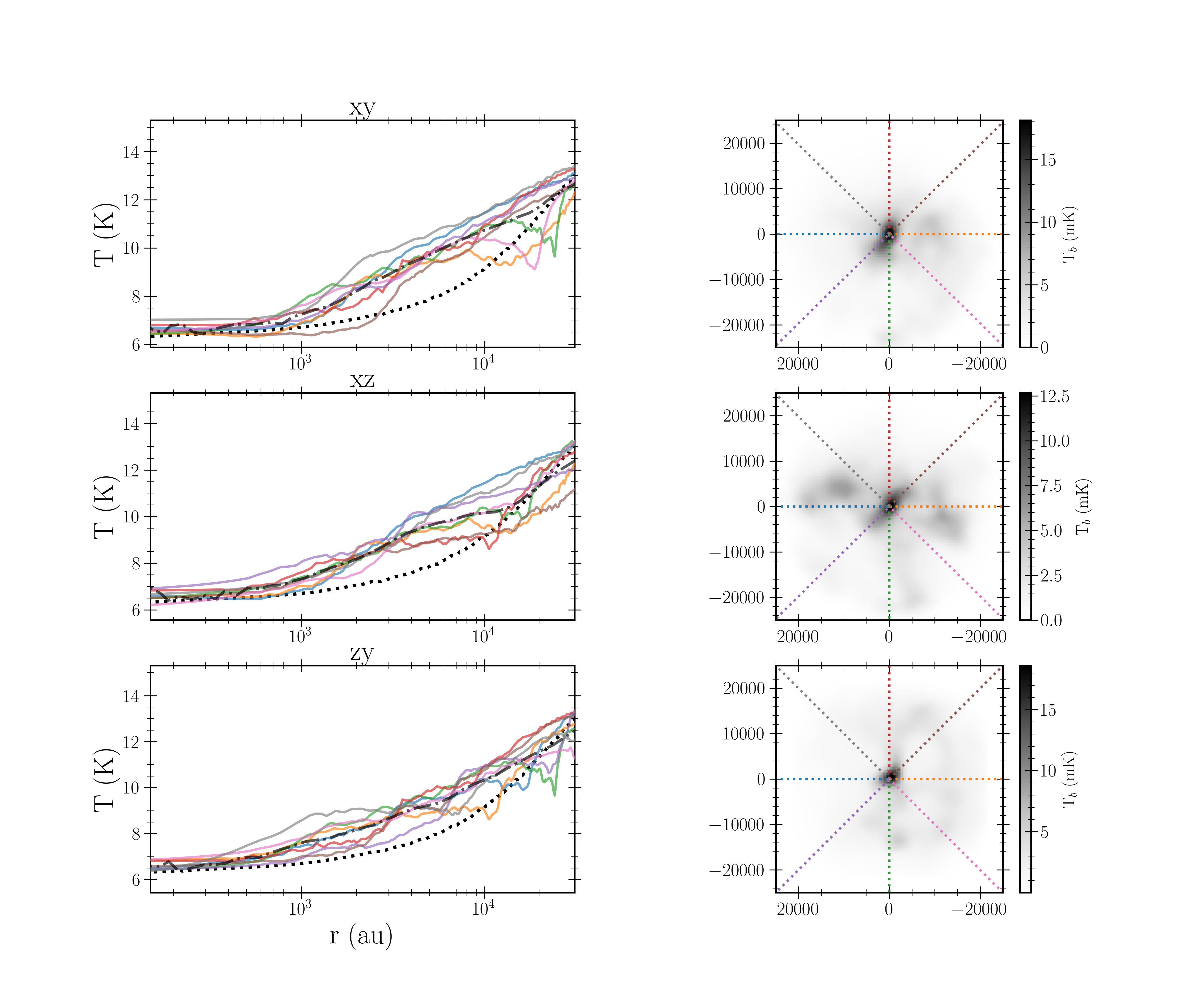}}
  \caption{The thermal structure of the simulated core. \emph{Left}: Radial profiles for the dust temperature along 8 different radial cuts on the x-y, z-x, and z-y planes. The profiles for the 1D hydrodynamical model from \citet{2010MNRAS.402.1625K} is indicated by the black dotted line.  \emph{Right}: The greyscale images shows the continuum emission of the core at 1.1~mm, while contours indicate 15\%, 30\%, 45\%, 60\%, and 75 \% of the peak intensity. The dotted colored lines indicate the direction along which the radial profiles in the other panel were extracted.}
     \label{fig:radial_td}
\end{figure*}   

\begin{figure*}[ht]
\resizebox{\hsize}{!}
        {\includegraphics{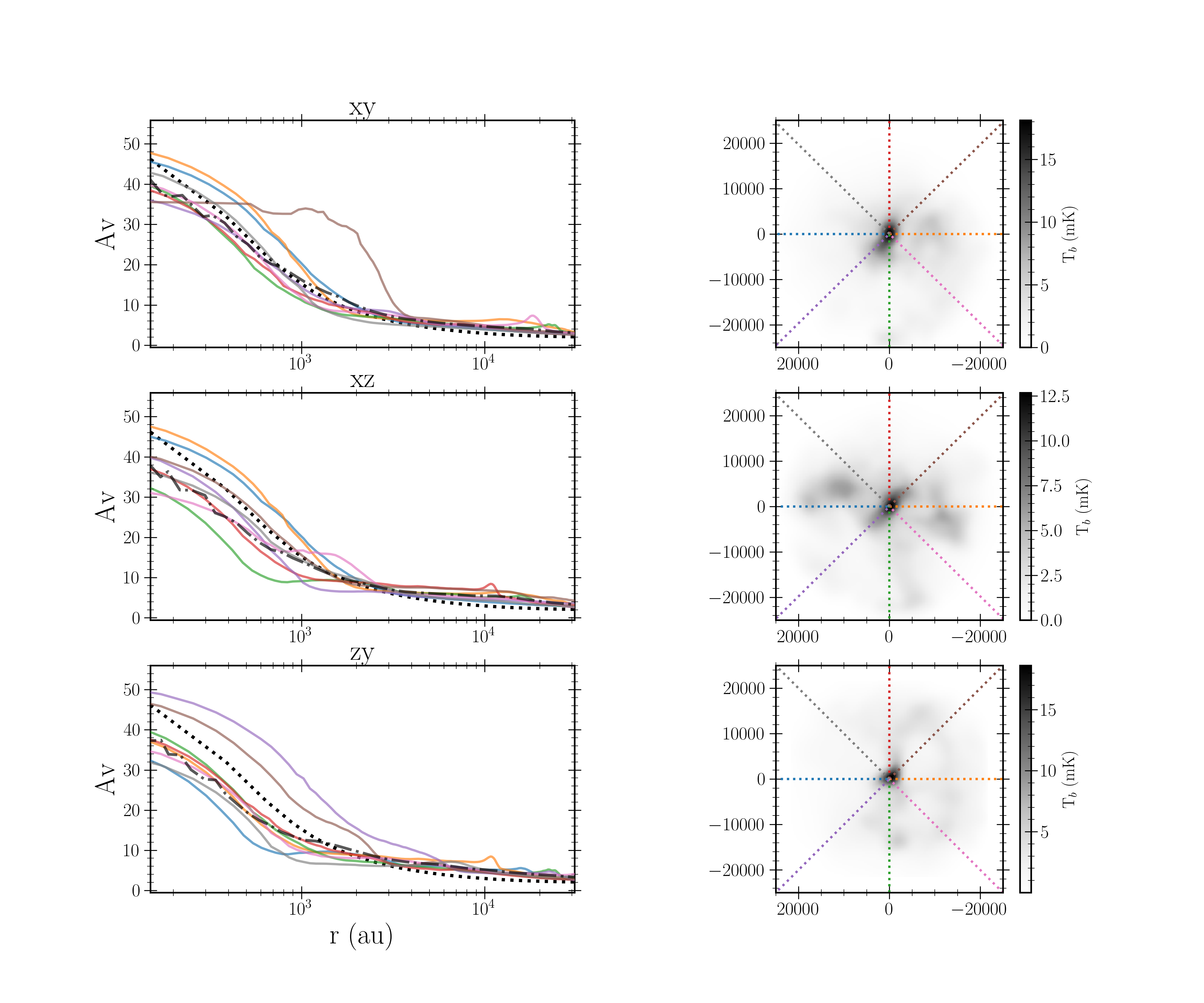}}
  \caption{The visual extinction within the simulated core. \emph{Left}: Radial profile for the visual extinction along 8 different radial cuts on the x-y, z-x, and z-y planes. The profiles for the 1D hydrodynamical model from \citet{2010MNRAS.402.1625K} is indicated by the black dotted line.  \emph{Right}: The greyscale images shows the continuum image of the core at 1.1~mm, while contours indicate 15\%, 30\%, 45\%, 60\%, and 75 \% of the peak intensity. The dashed colored lines indicate the direction along which the radial profiles in the left panel were extracted.}
     \label{fig:radial_av}
\end{figure*}

\begin{figure*}[ht]
\resizebox{\hsize}{!}
        {\includegraphics{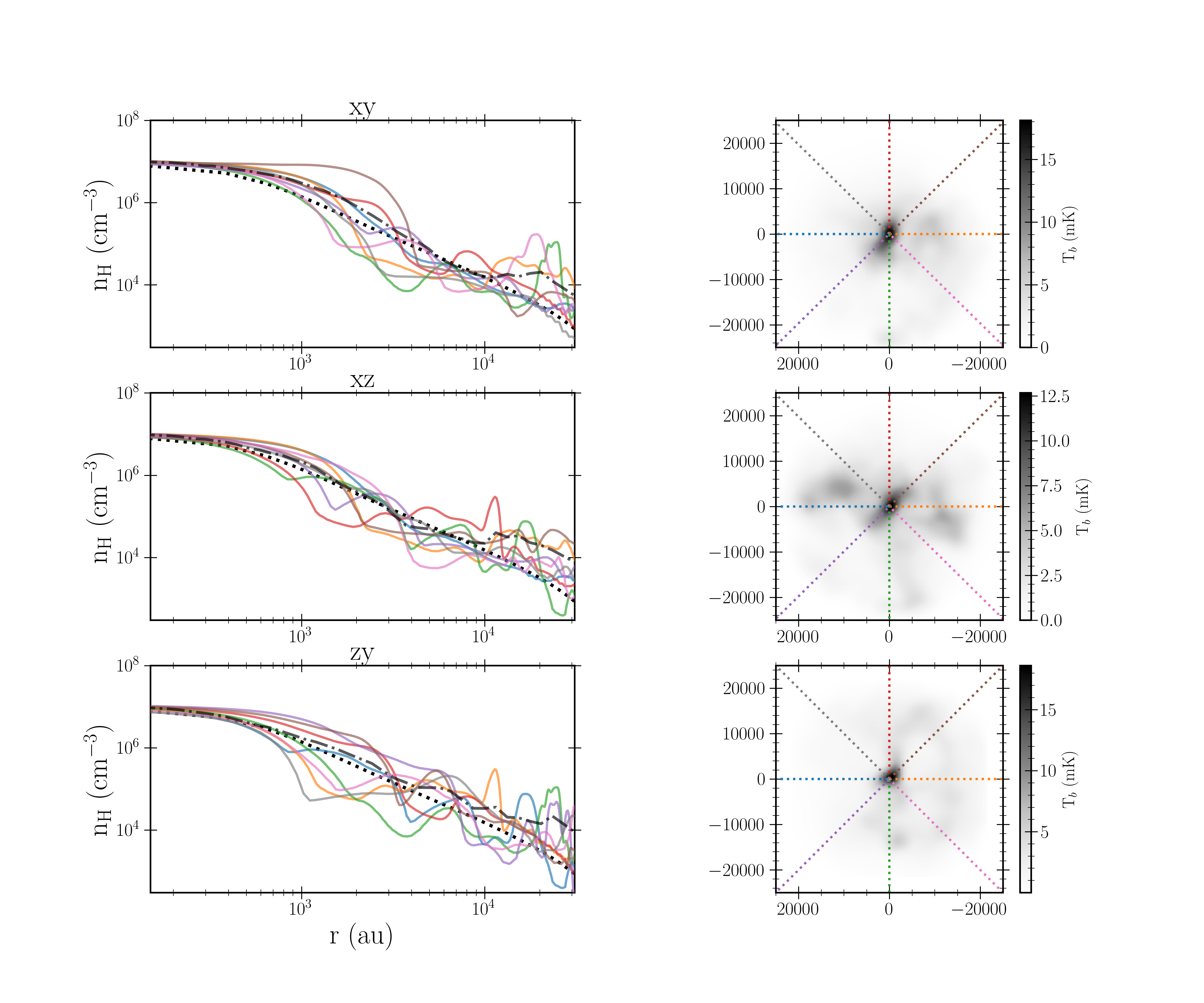}}
  \caption{The density structure of the simulated core. \emph{Left}: Radial profile for the number density along 8 different radial cuts on the x-y, z-x, and z-y planes. The profiles for the 1D hydrodynamical model from \citet{2010MNRAS.402.1625K} is indicated by the black dotted line.  \emph{Right}: The greyscale images shows the continuum image of the core at 1.1~mm, while contours indicate 15\%, 30\%, 45\%, 60\%, and 75 \% of the peak intensity. The dashed colored lines indicate the direction along which the radial profiles in the left panel were extracted.}
     \label{fig:radial_dd}
\end{figure*}   
\FloatBarrier

\section{Kernel density estimates.}\label{app:kde}
\begin{figure}[ht]
\resizebox{\hsize}{!}
        {\includegraphics{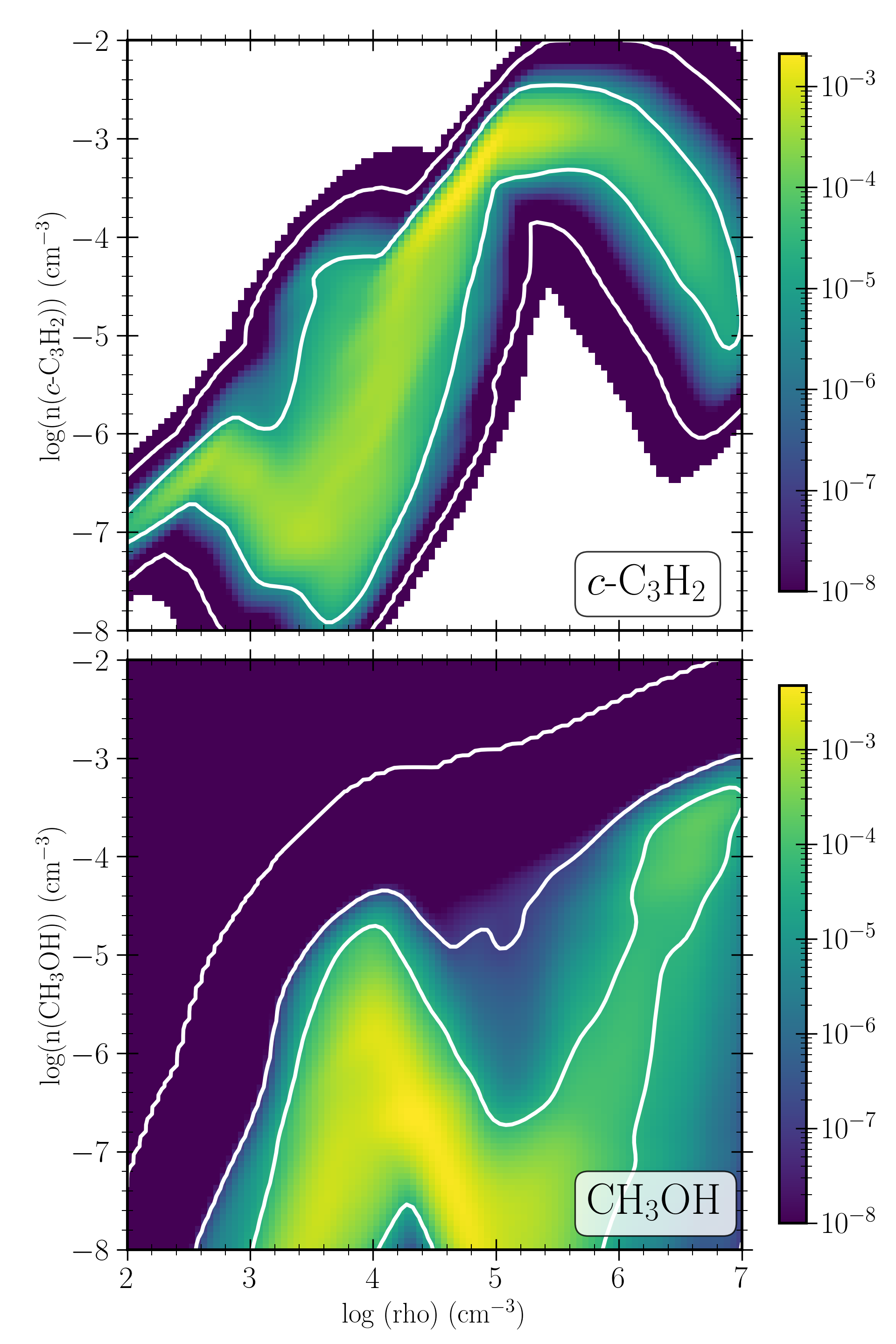}}
  \caption{Gaussian KDE for the absolute chemical abundances as a function of number density for each cell in the 3D model. Contours show the 25th, 50th, and 75th percentiles. Top panel shows $c$-C$_3$H$_2$ and bottom panel shows CH$_3$OH.}
     \label{fig:KDE_rho}
\end{figure}

\begin{figure}[ht]
\resizebox{\hsize}{!}
        {\includegraphics{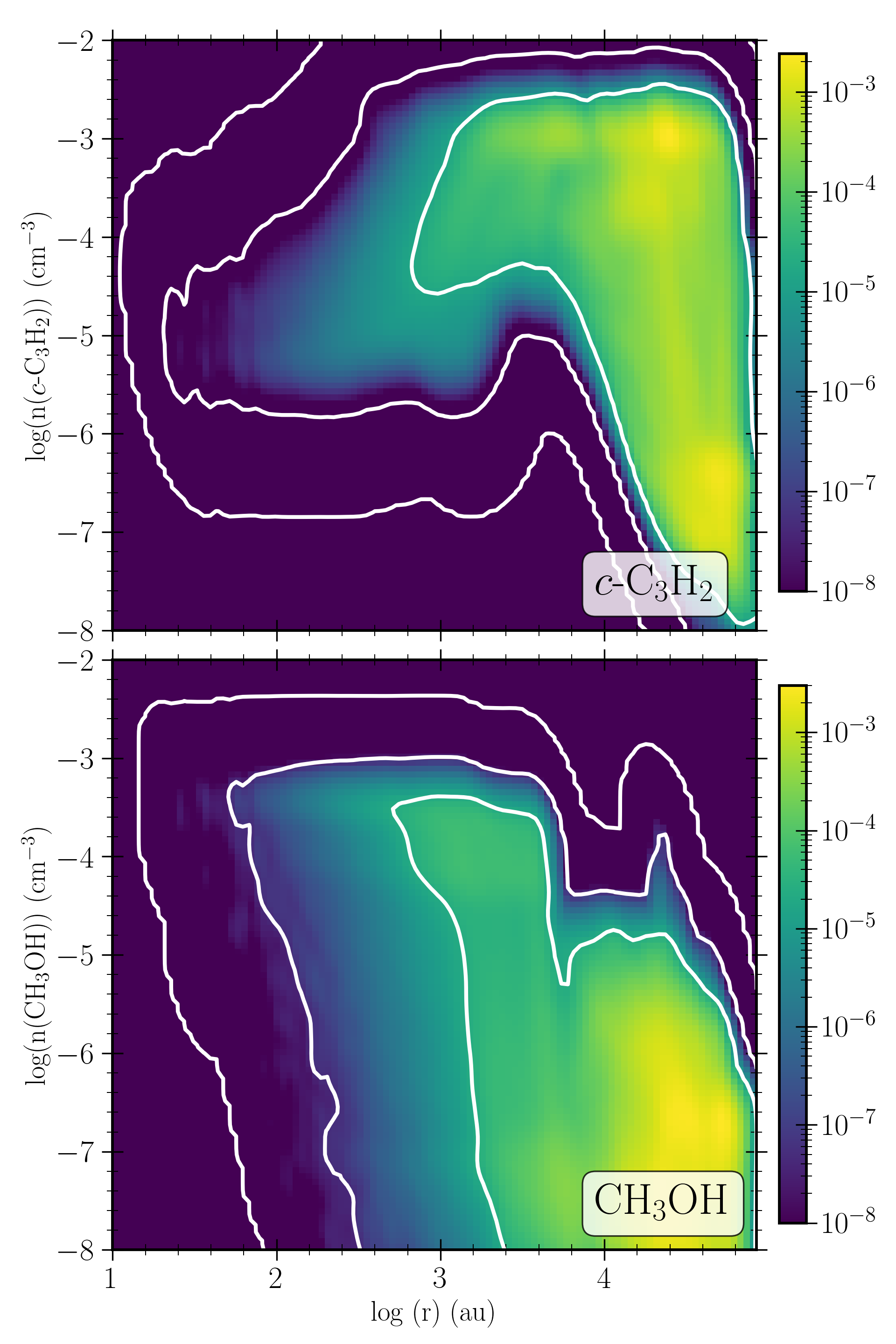}}
  \caption{Gaussian KDE for the absolute chemical abundances as a function of radius for each cell in the 3D model. Contours show the 25th, 50th, and 75th percentiles. Top panel shows $c$-C$_3$H$_2$ and bottom panel shows CH$_3$OH.}
     \label{fig:KDE_radius}
\end{figure}

\begin{figure}[ht]
\resizebox{\hsize}{!}
        {\includegraphics{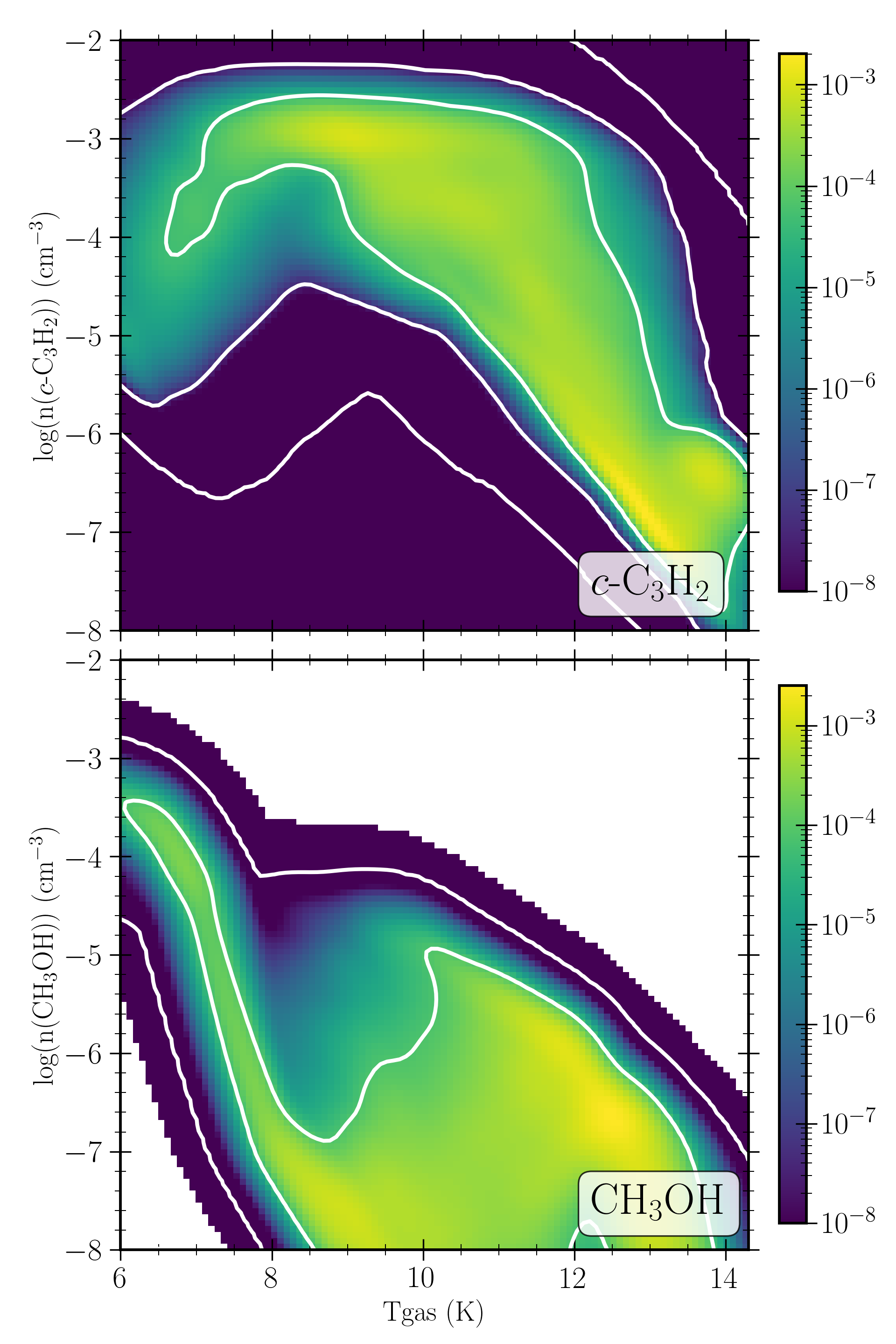}}
  \caption{Gaussian KDE for the absolute chemical abundances as a function of temperature for each cell in the 3D model. Contours show the 25th, 50th, and 75th percentiles. Top panel shows $c$-C$_3$H$_2$ and bottom panel shows CH$_3$OH. For CH$_3$OH, the 25th percentile is not visible is it is below the limit of the colorbar.}
     \label{fig:KDE_tgas}
\end{figure}

\FloatBarrier
Figures \ref{fig:KDE_rho} and \ref{fig:KDE_radius} show the Gaussian KDE for $c$-C$_3$H$_2$ and CH$_3$OH as a function of density and radius. Figure \ref{fig:KDE_tgas} shows the Gaussian KDE as a function of temperature.

\FloatBarrier

\onecolumn
\section{Model comparison for plane z-x and z-y.}\label{app:3}

\begin{figure*}[ht]
\resizebox{\hsize}{!}
        {\includegraphics{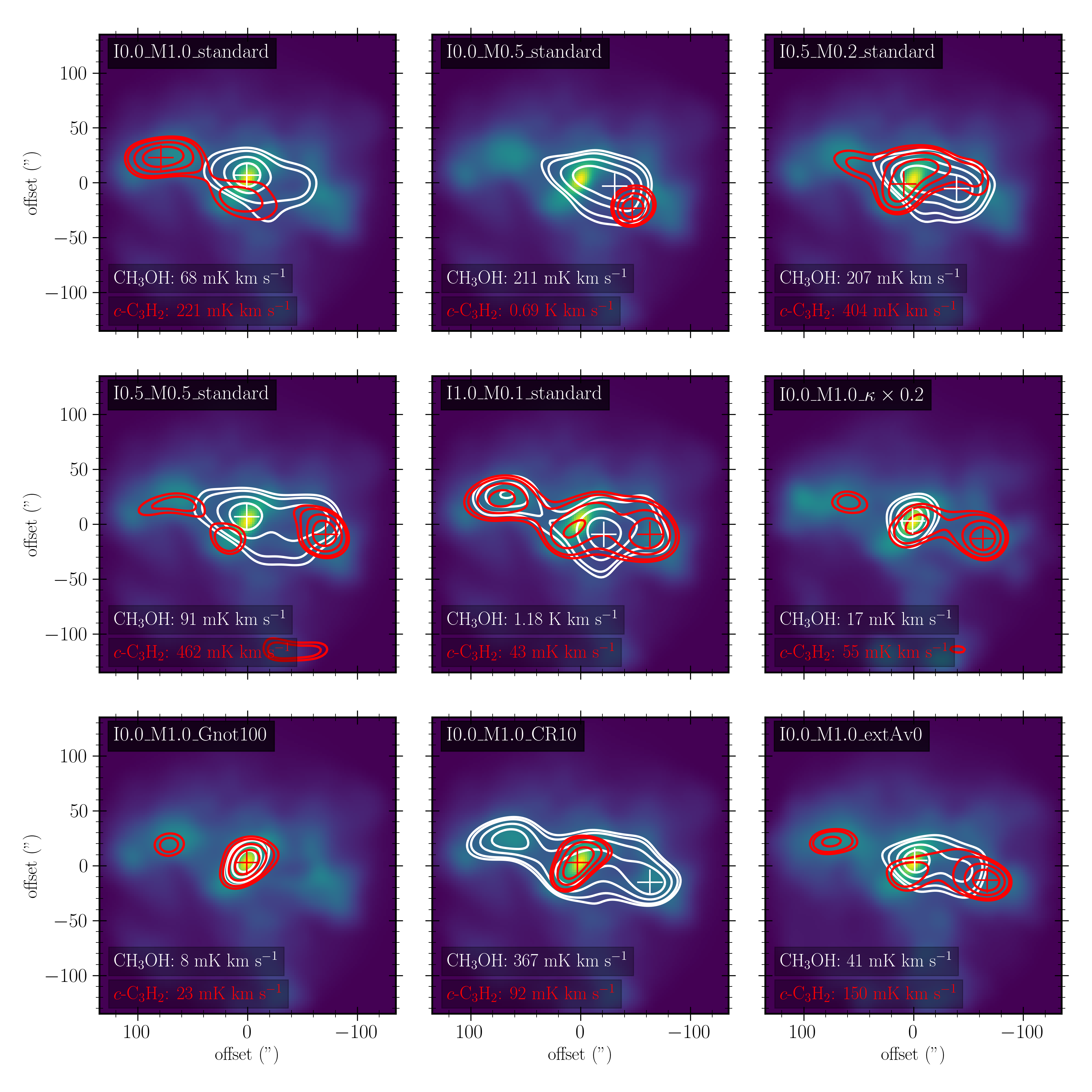}}
  \caption{Comparison between the $c$-C$_3$H$_2$ 3$_{2,2}$--3$_{1,3}$ transition (red contours) and CH$_3$OH 2$_{1,2}$--1$_{1,1}$ (white contours) integrated emission for each of the 9 chemical models listed in Table \ref{tab:chemical_models}. The layout of the figure is similar to that of Fig. \ref{fig:model_comparison} but for the z-x plane in the model.}
     \label{fig:C3H2_CH3OH_comparison2}
\end{figure*}

\begin{figure*}[ht]
\resizebox{\hsize}{!}
        {\includegraphics{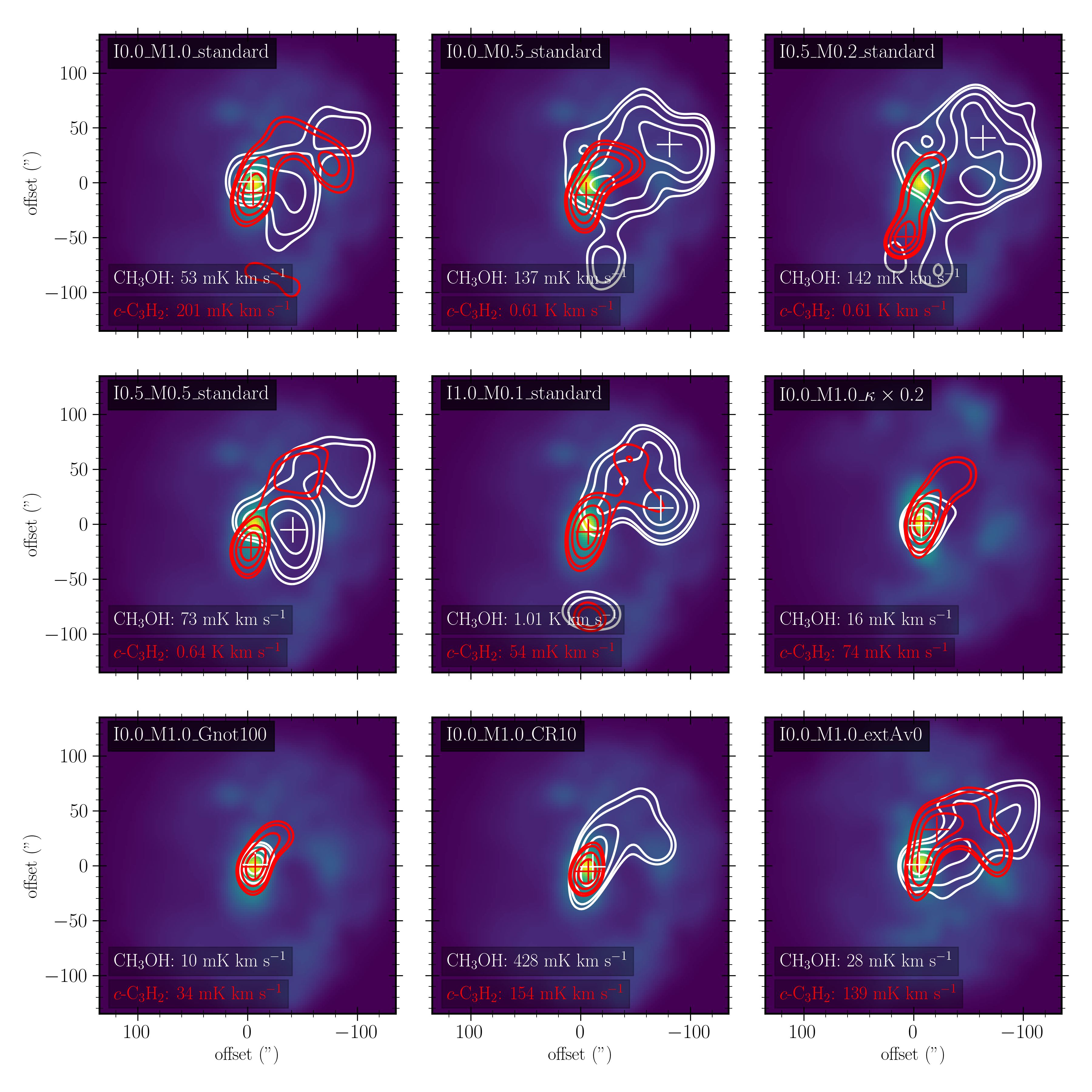}}
  \caption{Comparison between the $c$-C$_3$H$_2$ 3$_{2,2}$--3$_{1,3}$ transition (red contours) and CH$_3$OH 2$_{1,2}$--1$_{1,1}$ (white contours) integrated emission for each of the 9 chemical models listed in Table \ref{tab:chemical_models}. The layout of the figure is similar to that of Fig. \ref{fig:model_comparison} but for the z-y plane in the model.}
     \label{fig:C3H2_CH3OH_comparison3}
\end{figure*}   

\FloatBarrier
\clearpage

\end{appendix}
\end{document}